\documentclass[preprint,12pt]{elsarticle}

\usepackage{amssymb}

\usepackage{framed,multirow}

\usepackage{amsmath, amssymb}
\usepackage{latexsym}

\usepackage{url}
\usepackage{xcolor}

\usepackage{listings}
\usepackage{tensor}
\usepackage{tikz}
\usepackage{tikz-qtree}
\usepackage{caption}
\newcommand{\SpEC}{{\texttt{SpEC}}}

\definecolor{newcolor}{rgb}{.8,.349,.1}

\begin{document}

\begin{frontmatter}



\title{Automatic generation of CUDA code performing tensor manipulations using C++ expression templates}


\author[1,2,3]{Adam G.M. ~Lewis\corref{cor1}}
\cortext[cor1]{Corresponding author: 
  Tel.: +1-519-569-7600;  
  e-mail: alewis@perimeterinstitute.ca}
\author[1,4]{Harald P. ~Pfeiffer}

\address[1]{Canadian Institute for Theoretical Astrophysics, 60 St George St, Toronto, M5S 3H8, Ontario, Canada}
\address[2]{Department of Physics, University of Toronto, 60 St George St, Toronto, M5S 1A7, Ontario, Canada}
\address[3]{Perimeter Institute for Theoretical Physics, 31 Caroline St. N, Waterloo, N2L 2Y5, Ontario, Canada}
\address[4]{Max-Planck-Institut f\"{u}r Gravitationsphysik (Albert-Einstein-Institut), Wissenschaftspark Potsdam-Golm, Am M\"{u}hlenberg 1, 14476, Potsdam, Germany}

\begin{abstract}
  We present a C++ library, \texttt{TLoops}, which uses a hierarchy
  of expression templates to represent operations upon tensorial
  quantities in single lines of C++ code that resemble analytic equations.
  These expressions may be run as-is, but may also be used to emit equivalent
  low-level C or CUDA code, which either performs the operations more
  quickly on the CPU, or allows them to be rapidly ported to run on NVIDIA
  GPUs. We detail the expression template and C++-class hierarchy that
  represents the expressions and which makes automatic code-generation
  possible. We then present benchmarks of the expression-template code, 
  the automatically generated C code, and the automatically generated CUDA
  code running on several generations of NVIDIA GPU.
\end{abstract}


\end{frontmatter}


\section{Introduction}

Partial differential equations that involve vectorial or tensorial
quantities are very common in science. For example, the vacuum Maxwell's
equations,
\begin{eqnarray}\label{eq:Maxwell}
  \partial_t\vec E&=\nabla\times\vec B,\\
  \partial_t\vec B&=-\nabla\times\vec E, \\
\nabla \cdot \vec{B} &= 0, \\
\nabla \cdot \vec{E} &= 0,
\end{eqnarray}
involve manipulations of the vector fields $\vec E$ and $\vec B$. 
Numerically, these fields are represented by arrays of three numbers per
point on a discretized spatial grid. Manipulations of the fields using a language
such as C or FORTRAN will involve loops over each component and over the grid-size.

When solving Einstein's equations of general
relativity~\cite{baumgarteShapiroBook}, the necessary tensorial
equations can become quite involved.  As a moderate example
consider the evolution equation of the spatial metric in certain
formulations of Einstein's equations,
\begin{equation}\label{eq:dtg_ij}
  \partial_t g_{ij} = -2\alpha K_{ij} + \nabla_i\beta_j +\nabla_j\beta_i,\qquad i,j=1,2,3.
\end{equation}
Here, $g_{ij}$ and $K_{ij}$ are the spatial metric and the extrinsic
curvature; both these are represented by spatially varying, symmetric
$3$x$3$ matrices.  The scalar quantity $\alpha$ denotes the lapse-function and $\beta_i$ the
shift-vector, both of which are also spatially varying.  And finally,
$\nabla_i$ denotes the covariant derivative operator compatible with $g_{ij}$.  
Equations~(\ref{eq:Maxwell})--(\ref{eq:dtg_ij})
depend on one or two indices, respectively.  Intermediate expressions
in general relativity can easily depend on more indices, for instance
the Christoffel-symbols are defined as
\begin{equation}\label{eq:SpatialChristoffel}
  \Gamma^i_{jk} = \frac{1}{2}\sum_{l=1,2,3} g^{il}\left(\partial_kg_{jl}+\partial_jg_{lk}-\partial_lg_{jk}\right),\qquad i,j,k=1,2,3.
\end{equation}
where $\partial_i$ denotes the partial derivative, and the 3x3
symmetric matrix $g^{ij}$ is the inverse of the matrix $g_{ij}$, 
both of which are spatially varying.
Because of the symmetry in the index-pair $jk$,
Eq.~(\ref{eq:SpatialChristoffel}) represents 18 independent equations,
each one with nine terms on the right-hand side.

Henceforth, we adopt the Einstein sum-convention
that repeating indices are being summed over (i.e. we will no longer
write $\sum_l$ in equations like Eq.~(\ref{eq:SpatialChristoffel})).
Furthermore,  Latin lower-case letters
from the middle of the alphabet ($i,j,k,\ldots$) will range over the three
spatial dimensions.

Upon spatial discretization, each spatially dependent tensor is
represented on a spatial grid or, for multi-domain methods, 
on multiple spatial grids.  On each
such grid, assumed to have $N$ points, Eq.~(\ref{eq:dtg_ij}) would
then, schematically, be represented by code such as that in
Listing~\ref{lst:Example-Schematic}.

\begin{lstlisting}[caption={\label{lst:Example-Schematic}
      Schematic implementation of Eq.~(\ref{eq:dtg_ij})}, language=C]
  Tensor<DataMesh> dtg, K, db;
  DataMesh alpha;
  // initialize dtg, K, db, alpha
  for(int i=0; i<3; ++i) {
    for(int j=0; j<=i; ++j) {
      for(int a=0; a<N; ++a) {
        dtg(i,j)[a]=-2*alpha[a]*K(i,j)[a]+db(i,j)[a]+db(j,i)[a];
      }
    }
  }
\end{lstlisting}

The schematic listing~\ref{lst:Example-Schematic} indexes tensorial
objects with parentheses for the tensor-indices; the grid-points of
the underlying grid are indexed with square brackets.  We furthermore
assume in Listing~\ref{lst:Example-Schematic} that the covariant
derivative of $\beta_i$ was already precomputed\footnote{The covariant derivative is given by
  $$
  \nabla_j\beta_i = \partial_j\beta_i - \Gamma^k_{ij}\beta_k,
  $$
  where the last term in this expression uses the sum-convention.
}  into the variable
\verb$db$.

Our focus in this paper is the numerical relativity code \SpEC{},
\cite{SpEC},
a mature code in active use for the computation of gravitational
waveforms for ground-based detectors. Expressions such as
that of Listing~\ref{lst:Example-Schematic} are ubiquitous in
\SpEC{} and present a major challenge to development, adaptation,
and maintenance.  The library presented in this paper, \texttt{TLoops},
removes from \SpEC{} the need for explicit source-code loops over
tensor-indices.  Equation~(\ref{eq:dtg_ij}) can
then be written as a single line, as illustrated in
Listing~\ref{lst:Example-TLoops}.

\begin{lstlisting}[caption={\label{lst:Example-TLoops}Implementation of Eq.~(\ref{eq:dtg_ij}) in \SpEC{} with the implicit tensor-loop functionality presented in this paper.}, language=C++]
  Tensor<DataMesh> dtg, K, db;
  DataMesh alpha;
  // initialize dtg, K, db, alpha
  dtg(Sym<0,1>(), i_, j_) = -2*alpha*K(i_,j_)+db(i_,j_)+db(j_, i_);
\end{lstlisting}

The variables \verb$i_, j_$, etc, are pre-defined by \texttt{TLoops}. 
Overloaded indexing-operators and
assignment-operators are defined such that the single line in
Listing~\ref{lst:Example-TLoops} expands to \emph{all} relevant loops, both
over tensor-indices and over grid-points.
\texttt{TLoops} also handles sums. 
For instance Eq.~(\ref{eq:SpatialChristoffel}) can be coded as the 
single expression in listing~\ref{lst:Christoffel-TLoops}.

\begin{lstlisting}[caption={\label{lst:Christoffel-TLoops}
      Implementation of Eq.~(\ref{eq:SpatialChristoffel}) in \SpEC{} with the implicit tensor-loop functionality presented in this paper.}, language=C++]
  Tensor<DataMesh> Gamma, Invg;
  Tensor<Tensor<DataMesh>> dg; // partial_k g_{ij} = dg(i,j)(k)
  // initialize Gamma Invg, dg
  Gamma(Sym<1,2>(), i_, j_, k_) =
    0.5*Sum(l_, Invg(i_, l_)*(dg(j_,l_)(k_)+dg(l_,k_)(j_)-dg(j_,k_)(l_)));
\end{lstlisting}

There already exist several packages
implementing similar functionality \cite{FTensorpaper, dealII85, jeremicTensor, pooma, SVMT, BLITZ}. Consistent with our observations,
benchmarks of them show impaired performance relative
to explicitly coded loops \cite{FTensorbenchmarks}, presumably due to 
compiler optimizations being oriented towards the latter. 

The true (and to our knowledge unique) advantage
of our package is its ability to automatically generate equivalent 
source code to templated expressions. When a certain compiler
flag is defined, \texttt{TLoops} stores each unique tensor expression
it encounters within the linker code of each compiled library. A packaged
executable, \texttt{CodeWriter}, thus has access to the full list of 
possible tensor expressions, from which it generates legal \emph{non-templated} code
performing equivalent operations. We present here two examples, a (loop based)
C-implementation, and a GPU (CUDA) implementation. In either case, the original C++ template-code does
not need any source-code modifications -- the new C- or CUDA-code is incorporated
at link-time.

Because of the latter
functionality, \texttt{TLoops} can be used to immediately port large
numbers of tensor operations to the GPU without the need to explicitly
write kernels. These tensor operations are normally substantially faster
than CPU code, and allow data to be kept on the GPU
between calls to other GPU kernels, allowing segments of code to be hand-ported
without extraneous CPU-GPU synchronizations. 

The remainder of this paper is divided into three parts: First, we introduce
\SpEC{} and outline
the C++ template techniques that enable the compact code in
listings~\ref{lst:Example-TLoops} and~\ref{lst:Christoffel-TLoops}.
Second, we present our techniques to allow replacement of the
template-generated code with automatically generated non-templated code.
Finally, we show detailed benchmarks of the new results.

\section{Direct evaluation of tensor loops using expression templates}
\subsection{Spectral Einstein Code}

The code presented in this paper is based on the Spectral Einstein
Code (\SpEC)~\cite{SpEC} written in C++.  In
\SpEC, arrays over grid-points are represented by the
\verb$class DataMesh$, and tensorial objects are represented by \\
\verb$template<class T> class Tensor$.  \SpEC's
\verb$class DataMesh$ already contains expression-templates that
handle loops over grid-points.  Therefore, in \SpEC,
Listing~\ref{lst:Example-Schematic} is coded as displayed in Listing \ref{lst:Example-SpEC}.
\begin{lstlisting}[caption={\label{lst:Example-SpEC}Implementation of Eq.~(\ref{eq:dtg_ij}) in \SpEC.}, language=C++]
  Tensor<DataMesh> dtg, K, db;
  DataMesh alpha;
  // initialize dtg, K, db, alpha
  for(int i=0; i<3; ++i) {
    for(int j=0; j<=i; ++j) {
      dtg(i,j) = -2*alpha*K(i,j) + db(i,j) + db(j,i);
    }
  }
\end{lstlisting}

Indexing a \verb$Tensor<T>$, e.g. \verb$K(i,j)$, returns a (const or
non-const) reference to \verb$T$.
\SpEC's \verb$class Tensor$ is aware of symmetric indices.  For
instance, if \verb$K$ is initialized as symmetric, \verb$K(i,j)$ and
\verb$K(j,i)$ both return a reference to the \emph{same} element.

\SpEC's \verb$class DataMesh$ implements automatic resizing when assigned
to.  Furthermore, when used on the right-hand-side of assignments (as
in Listing~\ref{lst:Example-SpEC}), \verb$DataMesh$ checks consistency
of the sizes of all \verb$DataMesh$'es involved.  These consistency
checks, combined with the absence of the explicit loop over
grid-points and indexing of grid-points already significantly reduces
the possibility of coding errors.  However, two major shortcomings remain:
\begin{enumerate}
\item Loops over tensor indices must be coded manually, which is 
  tedious and error-prone. Specifically, the
  loops over indices must be consistent with the symmetries of
  the respective \verb$Tensor$ (cf. the loop over \verb$j$ in
  Listing~\ref{lst:Example-SpEC}).
  \item The existing \SpEC{} expression templates operate on
    \verb$class DataMesh$'es. For each combination \verb$(i,j)$, the
    inner loop thus represents an independent expression on
    \verb$DataMesh$, triggering a
    full traversal of all gridpoints.  In Listing~\ref{lst:Example-SpEC} this requires six
    traversals of the associated memory, whereas Eq.~(\ref{eq:SpatialChristoffel}) 
    would require 18 traversals.
\end{enumerate}

\texttt{TLoops} corrects both these shortcomings by removing the need
to write explicit loops entirely. This is done using a hierarchy of 
expression templates to represent tensorial manipulations. In the
immediately following sections we detail the specifics of those templates. 

\subsection{Tensors and \textit{SpEC}'s \textit{Tensor} class}

For our purposes, tensors are objects with $R$ indices, each 
taking $D$ distinct values.  The integer
$R\ge 0$ is called the \emph{rank} of the tensor, and $D$ its
\emph{dimension}.  For instance, $g_{ij} (i,j=0,1,2)$ indicates a 
rank $R=2$ tensor of dimension $D=3$.
If a tensor is \emph{symmetric} on a pair of indices, then the
ordering of the two indices in the pair is irrelevant.  For instance,
$\Gamma^{i}_{jk}$ defined in Eq.~(\ref{eq:SpatialChristoffel}) is
symmetric on its two lower indices, i.e. $\Gamma^i_{jk}=\Gamma^i_{kj}$
for any values of $i, j, k=0, \ldots, D-1$.  The significance of the
index-placement (up/down) is irrelevant for the purposes of this
paper.  Many tensors in general relativity are symmetric on some or all 
indices, including $g_{ij}$ and $K_{ij}$ in
Eq.~(\ref{eq:dtg_ij})\footnote{Tensors can also be anti-symmetric, a
  property not implemented in \SpEC{} and therefore of no relevance
  here.}. In differential
  geometry tensors must satisfy additional conditions related to
  coordinate transformations, which are not satisfied by Christoffel
  symbols $\Gamma^i_{jk}$.  While $\Gamma^i_{jk}$ are not tensors in
  the mathematical sense, they are nevertheless represented in \SpEC{}
  with \texttt{class Tensor}.  

In general relativity one commonly encounters both
space-time and spatial tensors.  Indices of
space-time tensors range over the three spatial dimensions \emph{and}
time.  An example is the space-time metric,
\begin{equation}
  \psi_{ab},\qquad a,b=0,1,2,3,
  \end{equation}
where we use letters from the start of the
alphabet ($a, b, c \ldots$) to indicate space-time indices.  The zero-th index-value
(e.g. $a=0$) indicates the time-dimension, while $a=1,2,3$ indicate
the space dimensions.  The spatial metric $g_{ij}$ is a subset of the
space-time metric,
\begin{equation}\label{eq:g_from_psi_symbolic}
  g_{ij} = \psi_{(i+1)(j+1)}, \qquad i,j=0,1,2.
\end{equation}
Because \SpEC{} always indexes starting with $0$,
Eq.~(\ref{eq:g_from_psi_symbolic}) must add $1$ to the spatial indices
to obtain the relevant components of $\psi_{ab}$.

\SpEC's \texttt{Tensor} class represents multi-index objects whose
indices each take on the values $0,1,\ldots D-1$.  The represented
objects may be symmetric on some of their indices, like $g_{ij}$ or
$\Gamma^i_{jk}$.  Internally, a \texttt{Tensor<X>} holds an array of
elements, each an object \texttt{X}, of appropriate size given the symmetries
(e.g. the symmetric $D=3$ tensor $g_{ij}$ has six elements).  A
\texttt{Tensor<X>} furthermore holds a look-up table to translate
indices \texttt{(i,j)} into the actual storage location inside the
array.  Symmetries are implemented by the lookup table for
\texttt{(i,j)} and \texttt{(j,i)} pointing to the \emph{same} element.
\texttt{Tensor} is indexed with
parentheses, i.e. \texttt{Gamma(0,1,2)} represents $\Gamma^0_{12}$.
Listing~\ref{lst:SpEC-Tensor-Examples} demonstrates some
indexing-operations performed on \texttt{Tensor}, while 
Listing~\ref{lst:Example-SpEC} already demonstrated actual
computations performed with \texttt{Tensor}.  

\begin{lstlisting}[caption={\label{lst:SpEC-Tensor-Examples}
      Some typical \texttt{Tensor}-operations in \SpEC.},
    language=C++]
  Tensor<DataMesh> g, psi, beta; 
  // initialize g and beta with D=3, and psi with D=4.
  // Rank and symmetries as in main-text

  const int D=g.Dim();
  for(int i=0; i<3; ++i) {
    for(int j=0; j<=i; ++j) {
      g(i,j)=psi(i+1, j+1); //(*)
    }
  }

  for(int i=0; i<D; ++i) {
    beta(i) = psi(i+1, 0);
  }
\end{lstlisting}

The listings~\ref{lst:Example-SpEC} and \ref{lst:SpEC-Tensor-Examples}
use \texttt{class DataMesh}, another \SpEC{}-specific class.
\texttt{DataMesh} represents a multi-dimensional rectangular array,
holding one double per grid-point,
with dimension $D\ge 1$, extents $(N_0, N_1, \ldots N_{D-1})$ and size
$N=N_0\,N_1\cdots N_{D-1}$. 
\SpEC{}\_ implements expression templates: arithmetic operators between
\texttt{DataMesh}-objects and/or \texttt{double}'s are overloaded to
return recursively defined types encoding the operation and the 
data type of the operands (\texttt{DataMesh} or \texttt{double}).
The instantiations of the 
expression templates furthermore collect references to the memory locations of all
involved data.  The assignment operator then recurses through the 
template to evaluate the expression.

Certain design choices of \SpEC{} present challenges for the development
of \texttt{TLoops}. Because \SpEC{} is a well-established and
intensely used code, these choices cannot be changed and we
have to work within them:
\begin{itemize}
\item Dimension, rank and symmetry of a \texttt{Tensor<X>} are
  assigned dynamically at run-time, and not statically through
  template-arguments at compile-time.  This gives flexibility when 
  using instances of \texttt{Tensor}, because dimension/rank/symmetry can
  be changed as needed.  Unfortunately, this also implies that
  dimension/rank/symmetry are not available to C++'s type-system at
  compile-time.  Part of this paper therefore deals with injecting
  compile-time information into the
  tensor-expressions
  (e.g. Listings~\ref{lst:Example-TLoops}--\ref{lst:SpEC-Tensor-Examples})
  so that the information needed to construct loops over tensor-indices can be
  deduced at compile-time.
\item \texttt{Tensor<T>} is a template class which stores an array of
  \texttt{T}'s.  Because each \texttt{DataMesh} allocates its own
  storage independently, this implies that \texttt{Tensor<DataMesh>}
  has independent \texttt{double*} arrays of size $N$ for each
  tensor-component, rather than one contiguous array of size $N_{\rm
    components}\times N$.  While \SpEC{}'s design-choice makes it
  convenient to use \texttt{Tensor}s, it is not necessarily
  computationally optimal.  Specifically, for GPU-implementations of
  tensor-loops, the increased number of memory locations degrades
  performance, cf.~Section \ref{Benchmarks}.

  \item Finally, because of \SpEC{}'s age, and the need to run on
    various supercomputers with varying degree of up-to-date
    compilers, \SpEC{} restricts itself to C++03 with only a small 
    set of newer C++11 features, and no C++17-specific features.  
 \end{itemize}

\subsection{Capturing a TLoops-expression as a type}

\subsubsection{Classes for indexing a Tensor}

All expression templates begin with capturing the structure of the
expression as a \emph{type}.  Types are available to the compiler and
thus enable meta-programming at compile-time. In this section we
detail the classes we have developed to accomplish this.

Two classes represent indexing with tensor-indices, i.e. the variables
$i, j, \ldots$ appearing in tensorial equations like
Eqs.~(\ref{eq:dtg_ij}) and~(\ref{eq:g_from_psi_symbolic}):
\texttt{class TIndex} represents an index (e.g. $i$) across the entire
expression, whereas \texttt{class TIndexSlot} is specific to each
occurrence of $i$. 
\begin{lstlisting}[caption={\label{lst:TIndex-TIndexSlot}
      Classes \texttt{TIndex} and \texttt{TIndexSlot}.},
    language=C++]
  template<int dim, int label>
  class TIndex {
  public:
    static int Value() const { return mValue; }
    static bool Done() const { return mValue >= dim; }
    static bool Increment() const { ++mValue; }
    static bool Reset(const int ctr=0) { mValue=ctr; }
  private:
    static int mValue;
  };

  template<int dim, int label, int offset> 
  class TIndexSlot: public TIndex<dim, label> {
    public:
      static int Value() { return TIndex<dim,label>::Value()+offset; }
      static const int Offset=offset;
  };

  // define variables for use when coding tensor-loop expressions

  // 3-dimensional indices
  extern TIndexSlot<3,0,0> i_;
  extern TIndexSlot<3,1,0> j_;
  extern TIndexSlot<3,2,0> k_;
  extern TIndexSlot<3,3,0> l_;

  // 4-dimensional indices
  extern TIndexSlot<4,0,0> a_;
  extern TIndexSlot<4,1,0> b_;
  extern TIndexSlot<4,2,0> c_;
\end{lstlisting}

 Listing~\ref{lst:TIndex-TIndexSlot} defines two classes and a set of
 variables. \texttt{class TIndex<dim, label>} serves as a
 marker to enable type-capture.  As such, \texttt{TIndex} is
 templated on the dimension of the index. The additional
 \texttt{label}-argument distinguishes 
 indices of the same dimension (e.g. $i,j,\ldots$). \texttt{TIndex}
 also contains a counter ``mValue" and functionality to iterate 
 this counter over the allowed index-values of the given \texttt{TIndex}.
 This functionality will
 become relevant when we discuss evaluation of a tensor-expression in
 Section~\ref{sec:TLoops-eval}.

 \texttt{class TIndexSlot<dim, label, offset>} tags each individual occurance of an 
index in an expression.  This is required because different occurrances 
can have different offsets
 (i.e. ``$i$'' and ``$i+1$''), which therefore require different encodings.
 \texttt{TIndexSlot} inherits from \texttt{TIndex}, in order to allow
 easy down-casting, which is convenient for compile-time consistency
 checks of the index structure.

Finally, Listing~\ref{lst:TIndex-TIndexSlot} defines 
variables \texttt{i\_, j\_,} etc, that map to specific \texttt{TIndexSlots}.
These enable the user to inject type-information about indexing into source code,
conveniently resembling mathematical expressions in tensor calculus. 


\subsubsection{Types representing an indexed Tensor}

The classes and variables defined in the previous subsection
(\texttt{TIndex, TIndexSlot, i\_, ...}) are used to index a tensor,
e.g.  \texttt{g(i\_, 1)}.  This is accomplished with a suitable
\texttt{Tensor::operator()}, which will return a type that carries
all information about the indexing. This return-type is built from helper
classes introduced schmatically in Listing~\ref{lst:TIndexStructure}.
The first two helper classes (\texttt{TInequality} and \texttt{TSymmetry})
handle symmetries of tensors.

The marker-class \texttt{TInequality<pos1,pos2>} indicates a symmetry 
between \emph{one} pair of indices
in a tensor, namely the indices at \texttt{pos1}'th and
\texttt{pos2}'th position (where the counting starts from zero).  For
example, consider the metric $g_{ij}$ and its time-derivative
$\partial_tg_{ij}$ both of which are symmetric in their only two
indices.  When iterating over all components of these tensors, this
implies a condition $j\le i$, cf. Listing~\ref{lst:Example-SpEC}.
\texttt{TInequality<0,1>} precisely represents this inequality, in
that it indicates that in loops, the value of the 0-th index should be
equal or larger than the value of the 1-th index.

A generic tensor with any structure of symmetric
indices\footnote{Recall that we do not consider tensors with
  anti-symmetric pairs of indices, or with cyclic symmetries like those of the Riemann tensor.} 
can then be represented by a set
of \texttt{TInequality}'s. For a tensor without symmetries, this set
is empty.  For a tensor with one symmetric pair of indices, the set
contains one \texttt{TInequality}, like for $g_{ij}$ or for
$\Gamma^i_{jk}$ (\texttt{TInequality<1,2>}).  More generic symmetries
are represented by multiple \texttt{TInequality}'s.  For instance, a
rank-four tensor $A_{(ij)(kl)}$ separately symmetric on the first two
and last two indices is represented by
\texttt{TSymmetry<TInequality<0,1>, TInequality<2,3>>}, and a
completely symmetric rank-three tensor $B_{(ijk)}$ by
\texttt{TSymmetry<TInequality<0,1>, TInequality<1,2>>}.

Such sets of inequalities are represented by \texttt{class
  TSymmetry<...>}, which is only defined in specializations for
\texttt{TInequality<.,.>}, and which utilizes compile-time asserts to
enforce a monotonically increasing ordering of the \texttt{TInequality}'s.

Creation of the \texttt{TSymmetry<...>} marker-classes is handled via
free template functions \texttt{Sym<i,j>(), Sym<i,j,k>(), Sym<i,j,k,l>()} 
that return the \texttt{TSymmetry} class
representing full symmetrization on the indicated slots.  For the case
of several distinct symmetries, e.g. $A_{(ij)(kl)}$, \texttt{operator
    \&\&} is suitably overloaded to allow \texttt{Sym<0,1>() \&\&
    Sym<2,3>()}.

Knowledge of the symmetries of a tensor is only important on the
\emph{left-hand side} of an assignment, as only the indices on the left-hand 
side are looped over.  Therefore, it is optional to specify
symmetries for tensors on the right-hand side, as in
Listing~\ref{lst:Example-TLoops}. Doing so does not throw an error,
but also has no effect.  

\begin{lstlisting}[caption={\label{lst:TIndexStructure}
      Classes representing an indexed Tensor,
      i.e. its symmetries and how it was indexed.},
    language=C++]
  template<int pos1, int pos2> 
  struct TInequality {
    static_assert(pos1<pos2,
                  "TInequality must satisfy pos1<pos2");
  };

  template<class ...TInequalities> struct TSymmetry;

  template<class TSymmetry_t, class ...indices>
  struct TIndexStructure;

\end{lstlisting}

With symmetries of a tensor handled by \texttt{TSymmetry}, we now turn
to the indexing of a tensor.  This is handled by \texttt{class
  TIndexStructure<Tsymmetry<...symm>, ...indices>}, where the
template-pack \texttt{indices} has a length equal to the number of
indices.  Each type in this template-pack is either a
\texttt{TIndexSlot} or an \texttt{int}.  The former indicates indexing
with an implicit index (as used in Listing~\ref{lst:Example-TLoops}),
whereas the latter case indicates the usual indexing by an integer (as
used in Listing~\ref{lst:Example-SpEC}).  Indexing with implicit
indices and integers can be mixed.  Consider, for example, a rank 3
tensor symmetric on the last two indices, which is indexed on the
first slow with the integer 1.  In mathematical notation, this is
represented by $\Gamma^1_{ij}$, which is mapped by \texttt{TLoops} to \texttt{Gamma(1,i\_, j\_)} with indexing structure
\texttt{TIndexStructure<TSymmetry<TInequality<1,2>>, 1, TIndexSlot<3,0,0>, TIndexSlot<3,1,0>>}.

The classes \texttt{struct TSymmetry<...TIneq>} and
\texttt{TIndexStructure<TStructure\_t>, ...indices>} are recursively
defined in the number of \texttt{TInequality}'s and tensor-indices.
The specializations for the empty case are trivial. One then adds
additional inequalities/indices at the front via variadic template
arguments.  Each specialization defines several member types and member-variables
that will be useful subsequently, and which are shown in Listing~\ref{lst:TIndexStructure-variables}.

\begin{lstlisting}[caption={\label{lst:TIndexStructure-variables}
      Member types and variables for \texttt{TSymmetry} and
      \texttt{TIndexStructure}, which will be used in the automatic
      generation of the implicit loops.}, language=C++]

  template<class ...TInequalities>
  struct TSymmetry {
    // TSymmetry for tensor with first slot removed
    typename Shift_t; 
  };

  template<class TSymmetry_t, class ...indices>
  struct TIndexStructure {

    // ==== ADMINISTRATION ====

    // rank of tensor (= # of free indices + # of indexed indices)
    static const int Rank;
    // number of free indices
    static const int NFree;
    // number of distinct free indices
    static const int NUniqueFree;
    // set of all free indices
    typename TIndexSet_t;
    // TIndexStructure with 1st slot removed
    typename BASE;

    // ==== ITERATION ====

    // increment to next set of free indices
    void operator++()
    // iteration over free indices complete?
    operator bool() const;
    // reset iteration over all free indices
    void Reset();

    // ==== ACCESS ====

    // retrieve value of first index
    int GetFirstIndex() const;
    // retrieve value of N-th index (zero-counted)
    template<int N>
    int GetNthIndex() const;
    // retrieve values of all indeces as a size=Rank vector
    void GetAllIndices(MyVector<int>& idx) const;
  };
\end{lstlisting}

The member types and variables of \texttt{TSymmetry} and
\texttt{TIndexStructure} are used in subsequent steps to implement
functionality.  For instance, when adding two indexed tensors, the
suitable \texttt{operator+} will \texttt{static\_assert} that both
indexed tensors have the same set of free indices.  This mirrors
mathematical meaning, where $g_{ij} + \beta_i\beta_j$ is correct,
whereas $g_{ij}+\beta_i\beta_k$ is erroneous.  

\subsubsection{Expression-tree for implicit tensor loop expressions}

\texttt{TIndexStructure} represents the full indexed structure of an
indexed tensor, i.e. the information needed to prepare at compile-time
(via metaprogramming) the necessary loops.  In order to execute the
operation, in addition, the memory locations of all
\texttt{Tensor<DataMesh>} instances are needed.  The memory locations
will be stored in a recursive expression-tree assembled with a
template class \texttt{iBinaryOp<L, Op, R>} taking three
template-parameters: The first template-parameters
\texttt{L} and \texttt{R} represent the left and right operands.
These operands could be of type \texttt{iBinaryOp} themselves, thus
enabling the recursion to represent nested expressions.  The
middle template-parameter \texttt{Op} represents the mathematical
operation to be performed.

The '\texttt{i}' in \texttt{iBinaryOp} indicates that the relevant
expression is indexed by at least one symbolic tensor-index
(\texttt{i\_, j\_, ...}).  This is important, because for tensorial
expressions, only a small subset of mathematical operations are
permissible: (i) Addition and subtraction, for which the
tensor-indices in both operands must agree; (ii) Multiplication; and
(iii) negation.  In contrast, \SpEC{} expression-templates operating
on \texttt{DataMesh} utilize a much larger set of mathematical
operators, including division, and a greatly enhanced set of unary
operators (like square-root and trigonometric functions).

There are several groups of partial specializations of
\texttt{iBinaryOp}, to enable its full functionality.  These
specializations are schematically indicated in
Listing~\ref{lst:iBinaryOp}. 

\begin{lstlisting}[caption={\label{lst:iBinaryOp}
      Schematic specializations of \texttt{iBinaryOp}.}, language=C++]

  // Set (1):  Two indexed expressions
  iBinaryOp<iBinaryOp,MultOp,iBinaryOp>
  iBinaryOp<iBinaryOp,AddOp,iBinaryOp>
  iBinaryOp<iBinaryOp,SubOp,iBinaryOp>
  iBinaryOp<EmptyType,negateOp,iBinaryOp>


  // (2a) One indexed expression and one double
  //     d*iOp, iOp*d, IOp/d
  iBinaryOp<double,MultOp,iBinaryOp>
  iBinaryOp<iBinaryOp,MultOp,double>
  iBinaryOp<iBinaryOp,DivOp,double>

  // (2b) One indexed expression and one DataMesh
  //     DM*iOp, iOp*DM, iOp/DM
  iBinaryOp<DataMesh,MultOp,iBinaryOp>
  iBinaryOp<iBinaryOp,MultOp,DataMesh>
  iBinaryOp<iBinaryOp,DivOp,DataMesh>

  // (2c) One indexed expression and one scalar-valued
  //     DataMesh expression
  //     BOp*iOp, iOp*BOp, iOp/BOp
  iBinaryOp<BinaryOp,MultOp,iBinaryOp>
  iBinaryOp<iBinaryOp,MultOp,BinaryOp>
  iBinaryOp<iBinaryOp,DivOp,BinaryOp>

  // (3) leaf-node:  one indexed Tensor<DataMesh>
  iBinaryOp<TIndexStructure<TSymmetry<Symm...>, Indices...>,
            EmptyType,DataMesh>
\end{lstlisting}

\begin{itemize}
  \item[Set (1)] contains the recursive operators that combine two
    indexed expressions together.  As explained above, mathematically
    there are only four allowed operators wich are represented by the
    marker-classes \texttt{AddOp, SubOp, MultOp} and \texttt{negateOp}.
    For instance the `\texttt{+}' operators in
    Listing~\ref{lst:Example-TLoops} are represented by
    \texttt{iBinaryOp}'s of set (1).

\texttt{iBinaryOp<EmptyType,negateOp,iBinaryOp>} illustrates our
convention to indicate unary operators with the type \texttt{class
  EmptyType \{ \}} in lieu of the first template argument \texttt{L}
to \texttt{iBinaryOp}.

    \item[Sets (2)] recursively combine an indexed expression with a
      scalar expression (either a \texttt{double}, or a
      \texttt{DataMesh}, or a (scalar-valued) expression template of
      \texttt{DataMesh}, represented by \texttt{class BinaryOp}.  Only
      multiplication and division is mathematically permissible,
      leaving only three cases each.  The multiplication
      \texttt{0.5*...} in listing~\ref{lst:Christoffel-TLoops} is
      represented by a specialization of set (2a), and the term
      \texttt{2*N*K(i\_, j\_)} is represented by a specialization of
      set (2c), combining the DataMesh-expression \texttt{2*N} with
      the indexed expression \texttt{K(i\_, j\_)}.

\item[Set (3)] is the entry point into the recursive
  \texttt{iBinaryOp}-representations; it represents one indexed
  \texttt{Tensor<DataMesh}.  Examples of this type include
  \texttt{K(i\_, j\_)} in listing~\ref{lst:Example-TLoops} as well as
  \texttt{psi(i\_+1, 0)} which arises when rewriting the last loop of
  listing~\ref{lst:SpEC-Tensor-Examples} in implicit tensor notation.
\end{itemize}

The construction of recursive \texttt{iBinaryOp}s is handled by the
relevant overloaded operators.  The recursive types of set (1) and (2)
are returned by suitably defined \texttt{operator+},
\texttt{operator-}, and \texttt{operator*}.
Leaf-nodes of set (3) are returned by suitably templated
\texttt{Tensor<DataMesh>::operator()}, which are provided in two
versions: with and without a first argument of type
\texttt{TSymmetry}.  Such a \texttt{TSymmetry} argument must be
provided on the \emph{left-hand-side} of assignments, like
\texttt{dg(Sym<0,1>(), i\_, j\_)} in Listing~\ref{lst:Example-TLoops},
    because the symmetry is required to determine the
    loop-bounds. The instance of \texttt{TSymmetry} passed into
    \texttt{Tensor<DataMesh>::operator()} is encoded in the templated
    return-type of this operator within the
    \texttt{TIndexStructure}-parameter inside the set (3) type in
    listing~\ref{lst:iBinaryOp}.  On the right-hand-side, compile-time
    information about the symmetry of the tensors is not needed and
    therefore, presently, it is optional to specify the symmetry
    through an extra first argument
    to \texttt{Tensor<DataMesh>::operator()}\footnote{All examples in
      this paper omit the symmetry specifiers on the right-hand-side.}.

All \texttt{iBinaryOp} specializations have certain member-types and
member-variables which are useful when assembling the types
recursively, and when evaluating the implicit tensor-loop expression.  These members are indicated in Listing~\ref{lst:iBinaryOp-members}.

\begin{lstlisting}[caption={\label{lst:iBinaryOp-members}
      Member types and variables of \texttt{iBinaryOp}.}, language=C++]
  class iBinaryOp<L, Op, R> {

    // member types
    using TIndexSet_t = ...;
    using ExpandIndices_t = ...;

    // member variables: references to sub-expressions
    const L& lhs;  // (absent for unary operators)
    const R& rhs;
  }
\end{lstlisting}

\texttt{TIndexSet\_t} is a template-type that represents the set of
all free indices; this set is used in \texttt{operator+} and
\texttt{operator-} to verify that both operands have the same free
indices.  \texttt{ExpandIndices\_t} is the
\texttt{DataMesh}-expression type that results when all implicit
tensor-indices are replaced by concrete values, i.e. when each
\texttt{Tensor<DataMesh>} is replaced by the \texttt{DataMesh} of one
of its components.  This type will be used when evaluating the
tensor-loop expression, cf. Sec.~\ref{sec:TLoops-eval}. Finally, the
references \texttt{lhs} and \texttt{rhs} are also needed during
evaluation of the tensor-loop expression, as they contain the concrete
memory locations of all relevant data.

\subsection{Evaluation of TLoops-template}
\label{sec:TLoops-eval}

The preceding sections describe the individual elements that
make up an implicit-tensor loop assignment, as in
Listing~\ref{lst:Example-TLoops}: The left-hand-side of this
expression expands to a \texttt{iBinaryOp} of Set(3), whereas the
right-hand-side expands to a \texttt{iBinaryOp} of arbitrary
complexity.  These two elements are combined via the member-assignment
operator \texttt{operator=()} of the left-hand-side's type.

\begin{lstlisting}[caption={\label{lst:iBinaryOp-assignment}
      Assignment operators for implicit tensor loops.}, language=C++]
template<class ...Symm, class ...Indices>
class iBinaryOp<TSymmetry<Symm...>, Indices...>,
                EmptyType,DataMesh> {

    // iBinaryOp on right-hand-side
    template<class L, class O, class R>
    void operator=(const iBinaryOp<L,O,R>& op) {
      CheckIndexEquality(op);
      CheckExtentsAndResizeToMatch(op);
      TLoopApply(*this, TSetEqualOp(), op);
    }

    // just a double on right-hand-side (e.g. to set to zero)
    void operator=(const double d) {
      TLoopApply(*this, TSetEqualOp(), d);
    }

    // DataMesh on right-hand-side   
    void operator=(const DataMesh& dm) {
      CheckExtentsAndResizeToMatch(dm);
      TLoopApply(*this, TSetEqualOp(), dm);
    }

    // repeat for +=, -=
    template<class L, class O, class R>
    void operator+=(const iBinaryOp<L,O,R>& op) {
      CheckIndexEquality(op);
      CheckExtentsAndResizeToMatch(op);
      TLoopApply(*this, TAddEqualOp(), op);
    }
    //  ... 
}
\end{lstlisting}

Listing~\ref{lst:iBinaryOp-assignment} indicates the structure of
these assignment operators.  There are several such assignment
operators depending on the type of operation (\texttt{=, +=, -=, *=,
  /=}) and depending on the right-hand-side type (\texttt{iBinaryOp},
\texttt{DataMesh}, \texttt{double}).  Only some combination of these
are mathematically permissible, and only those are defined.  As
appropriate, these operators check that free tensor-indices on the
left-hand-side and the right-hand-side match, and they resize the data
on the left-hand-side.  Then all these operators call a templated free
function \texttt{TLoopApply} for the actual computations.  This allows
us to handle the different types of assignment (\texttt{=, +=, -=, ...})
without code-duplication.

\begin{lstlisting}[caption={\label{lst:TLoopApply}
      Assignment of implicit tensor loops.}, language=C++]

  template<class L, class O, class ApplyOp,  class RHS>
  void TLoopApply(iBinaryOp<L,O,DataMesh>& lhs,
                  const ApplyOp&,
                  const RHS& rhs) {
    lhs.CheckUniqueIndices();
    lhs.CheckSymmetries();
    for(lhs.Reset(); lhs; ++lhs) {
      BinaryOpHolder<RHS> holder(rhs);
      ApplyOp::modify(lhs.ExpandIndices(), holder.op);
    }
  }
\end{lstlisting}

Listing~\ref{lst:TLoopApply} executes the actual calculations, and as such this listing requires detailed explanations:
\begin{enumerate}
\item \texttt{TLoopApply()} starts with safety checks:
  \texttt{CheckUniqueIndices} is a compile-time check that there are no
  repeated tensor-indices on the left-hand-side.  This test catches,
  for instance, the typo ``\texttt{dg(Sym<0,1>(), i\_, i\_)}" in the
    left-hand-side of Listing~\ref{lst:Example-TLoops}, which is
    mathematically forbidden.  \texttt{CheckSymmetries()} verifies
    that the stated symmetries in the assignment
    ---e.g. \texttt{Sym<0,1>()} in Listing~\ref{lst:Example-TLoops}---
    agree with the run-time symmetry-state of the respective tensor.
    Because of \SpEC{}'s design decision that symmetries of
    \texttt{Tensor<X>} are set at run-time, this test necessarily can
    only trigger run-time errors.

    \item The loop \texttt{for(lhs.Reset(); lhs; ++lhs)} forwards
      directly to the corresponding member-functions of
      \texttt{TIndexStructure} shown in
      Listing~\ref{lst:TIndexStructure-variables}.
      \texttt{TIndexStructure} uses recursive template-pack expansion
      to recurse through all tensor-indices.  The loop will modify the
      static member-variables \texttt{int TIndex<dim, label>::mValue}
      of the \texttt{TIndex}-types occuring on the left-hand-side,
      cf. Listing~\ref{lst:TIndex-TIndexSlot}.  During the
      \texttt{++lhs} increment, these variables will be
      reset whenever they reach this upper bound.  In this case, the \texttt{TInequality} parameters
      indicate the position of a potential other index with which an
      inequality (arising from a tensorial symmetry) must be
      satisfied.  If so, \texttt{int
        TIndexStructure::GetNthIndex<int>()} retrieves the current
      value of this other index, which is used in re-setting the index
      under consideration.  Overall, the assignment
      \texttt{dg(Sym<0,1>(), i\_, j\_)=...} in
      Listing~\ref{lst:Example-TLoops}, results in loops \\
      \texttt{for(j=0; j<3; ++j) \{ for(i=j; i<3; ++i) \{
        ... \} \} } \\
      where `\texttt{i}' represents \texttt{TIndex<3,0>::mValue} and
      `\texttt{j}' represents \texttt{TIndex<3,1>::mValue}.
    \item Inside the loop in Listing~\ref{lst:TLoopApply}, we must now 
      index each \texttt{Tensor<DataMesh>} on the right-hand-side
      `\texttt{rhs}' with the current set of index-values as stored
      inside the respective \texttt{TIndex<dim,label>::mValue}.  Upon
      such indexing, each \texttt{Tensor<DataMesh>} in the
      right-hand-side expression becomes a standard \SpEC{}\_
      \texttt{DataMesh}, and the expression-tree becomes a regular
      \texttt{DataMesh}-expression template tree of type
      \texttt{RHS::ExpandIndices\_t}
      (cf. Listing~\ref{lst:iBinaryOp-members}).  The helper-class
      \texttt{BinaryOpHolder} recursively descends through
      `\texttt{rhs}'s structure, and builds an instance of the
      \texttt{DataMesh}-expression with all data-references pointing
      to the appropriate elements of each \texttt{Tensor<DataMesh>}.
      \item Finally, the actual assignment happens in
        \texttt{ApplyOp::modify()}.  This member-function of the
        marker-classes \texttt{SetEqualOp, PlusEqualOp, MinusEqualOp}
        takes its second argument (i.e. the \texttt{DataMesh}
        expression template representation), and
        assigns/adds/subtracts it from its first argument (the
        \texttt{DataMesh} returned by indexing the
        \emph{left}-hand-side with the current set of tensor-indices),
        thus triggering execution of \SpEC{}\_ \texttt{DataMesh} expression
        template code.
\end{enumerate}

\subsection{Sum-operations}

Let us now turn to an exposition of contractions as in
Listing~\ref{lst:Christoffel-TLoops}.

The goal is to transform the expression \texttt{Sum(k\_, op[k\_])}
(schematically) into the expression \texttt{op[0]+op[1]+op[2]}.
Here `\texttt{op}' indicates a tensor-loops expression which may have
an arbitrary number of free indices. These should remain intact in the
output expression.  In our code, this is implemented with a
template-class \texttt{PartialSum<curr\_dim, TIndex\_t, iBinaryOp>}.
An instance of this class is responsible for handling the index-value
\texttt{curr\_dim} of the tensor-index \texttt{TIndex\_t}.  This class
recursively decrements \texttt{curr\_dim} via inheritance of
\texttt{PartialSum<curr\_dim-1, TIndex\_t, iBinaryOp>}, and during
recursion assembles the full sum.  The corresponding code is shown schematically in
Listing~\ref{lst:PartialSum}.

\begin{lstlisting}[caption={\label{lst:PartialSum}
      class \texttt{PartialSum} which forms the core of the implementation of \texttt{Sum}.}, language=C++]

  // Recursion:
  template<int curr_dim, class TIndex_t, class iBinaryOp_t>
  struct PartialSum:
    public PartialSum<curr_dim-1, TIndex_t, iBinaryOp_t> {

    using BASE=PartialSum<curr_dim-1, TIndex_t,iBinaryOp_t>;

    // (a) type of unrolled sum
    using ExpandIndices_t
      =BinaryOp<typename BASE::ExpandIndices_t,
                AddOp,
                typename iBinaryOp_t::ExpandIndices_t>;

    // (b) constructor creating the sum-expanded BinaryOpHolder
    PartialSum(const iBinaryOp_t& summand):
      BASE(summand),
      this_term( (TIndex_t::Reset(curr_dim), summand) ),
      partial_sum(BASE::partial_sum, this_term.op)
    { };

    const BinaryOpHolder<iBinaryOp_t> this_term;
    const ExpandIndices_t partial_sum;
  };


  // Break recursion (curr_dim=0)
  template<class TIndex_t, class iBinaryOp_t>
  struct PartialSum<0, TIndex_t, iBinaryOp_t> {
    using ExpandIndices_t=typename iBinaryOp_t::ExpandIndices_t;

    PartialSum(const iBinaryOp_t& summand):
      this_term( (TIndex_t::Reset(0), summand) ),
      partial_sum(this_term.op)
    { };

    const BinaryOpHolder<iBinaryOp_t> this_term;
    const ExpandIndices_t& partial_sum;
  };
\end{lstlisting}

In principle, the framework presented here could detect implicit sums
even without the explicit \texttt{Sum(...)}, by watching via
meta-programming for duplicate \texttt{TIndex<.,.>} in
\texttt{operator*}.  Walter Landry's FTensor behaves in this way and
presents the choice as a design feature \cite{FTensorpaper}. We choose not to implement such implicit loop
functionality for two reasons: First it would leave evaluation order
according to C++ operator precedence, and it is not guaranteed that
C++ precedence rules will result in optimal evaluation.  The
requirement to explicitly place \texttt{Sum(...)} in the code will
force the user to make an explicit choice of how terms will be
grouped, thus exhibiting the FLOPS implications more clearly.  Second,
sums exponentially increase the amount of FLOPs in an expression.  The
explicit occurrence of \texttt{Sum}, especially when repeated multiple
times in the same expression, acts as signal for potentially very
expensive operations.

\section{Automatic code generation}
\label{AutoGen}
In this section we describe \texttt{TLoops}' automatic code generation
functionality, which represents \texttt{TLoops} expressions with
equivalent C or CUDA code. 
First, \texttt{SpEC} is compiled with certain options what encode
each \texttt{TLoops} expression it uses. Next, an executable called
\texttt{CodeWriter} iterates through the encoded expressions and
outputs new code for each. \texttt{SpEC} is then recompiled with this
new code, which replaces the expression-templates described in Section 4.2
at link-time. This gives in total four different \texttt{SpEC} compilation
variants: \texttt{NonAccel}, as always; \texttt{CodeWriter}, to output
the new equivalent code; \texttt{AccelCPU}, to link in the automatically-generated
C code; and \texttt{AccelCUDA}, to link in the automatically-generated 
CUDA code.

Let us first give a high-level overview of the tools which generate this
code.
As described in Section 4.2, \texttt{TLoops} expressions are represented
as trees. Each node in the tree represents either an operator, in which
case it has either one or two subnodes, or actual data (of type 
\texttt{double}, \texttt{DataMesh}, or indexed \texttt{Tensor<DataMesh>}),
in which case it has no subnodes and we call it a ``leaf". 
The root node represents the type of assignment ($=$, $+=$, $-=$, or $*=$).
In Section 4.2 this expression tree is built at compile-time with
recursive templates, with the \emph{one} goal of executing the
encoded calculation.

\begin{figure}
\begin{center}
\hrulefill
$\partial_t g_{ij} = -2\alpha K_{ij} + \nabla_i \beta_j + \nabla_j \beta_i$
\hrulefill
\begin{verbatim} dg(Sym<0,1>(), i_, j_) = -2.*alpha*K(i_,j_) + beta(j_)(i_) + beta(i_)(j_); \end{verbatim}
\hrulefill
\begin{verbatim}
BOp<TIndStr<Sym<0, 1>, Ind<0>, Ind<1>>, EqualsOp,
  BOp<double, MultOp, 
     BOp<DataMesh, MultOp,
       BOp<TIndStr<Ind<0>, Ind<1>>, PlusOp,
         BOp<std::pair<TIndStr<Ind<1>>, TIndStr<Ind<0>>>, PlusOp, 
             std::pair<TIndStr<Ind<0>>, TIndStr<Ind<1>>>>>>>>;
\end{verbatim}

\end{center}

\hrulefill
\begin{verbatim}
for(int i=0; i<3; ++i){   
  for(int j=i; j<3; ++j){
    for(int x=0; x<GRIDSIZE; ++x){
\end{verbatim}
\begin{tikzpicture}
\Tree [ [.\texttt{TDm0[i][j][x] } ] [.\texttt{=}  
                [.\texttt{d0} ]  [.\texttt{*} 
                        [.\texttt{DM0[x]} ] [.\texttt{*} 
                           [.\texttt{TDm1[i][j][x]} ] [.\texttt{+}
                                [.\texttt{TTDm0[j][i][x]}  ] [.\texttt{+} 
                                        [.\texttt{TTDm1[i][j][x];} ] ] ]   ]  ] ] ]
\end{tikzpicture}
\begin{verbatim}
    }
  }
}
\end{verbatim}
\hrulefill
\caption{Various representations of the $\partial_t g_{ij}$ operation: in mathematical 
notation (top), as a \texttt{TLoops} expression (second), as a nested series
of \texttt{TLoops} expression templates (third), and as the automatically output code diagrammed with each fragment
in the appropriate part of the semantic tree used to represent expressions at runtime (bottom). We use shorthands for the class types
in the expression template: \texttt{BOp} for \texttt{iBinaryOp}, \texttt{TIndSt} for \texttt{TIndexStructure},
and \texttt{Ind} for \texttt{TIndexSlot}. Within the tree, \texttt{DM} represents \texttt{DataMesh} (i.e. a single component array),
\texttt{TDm} represents \texttt{Tensor<DataMesh>}, and \texttt{TTDm} represents \texttt{Tensor<Tensor<DataMesh>>}.}
\label{fig:kijtreefig}
\end{figure}
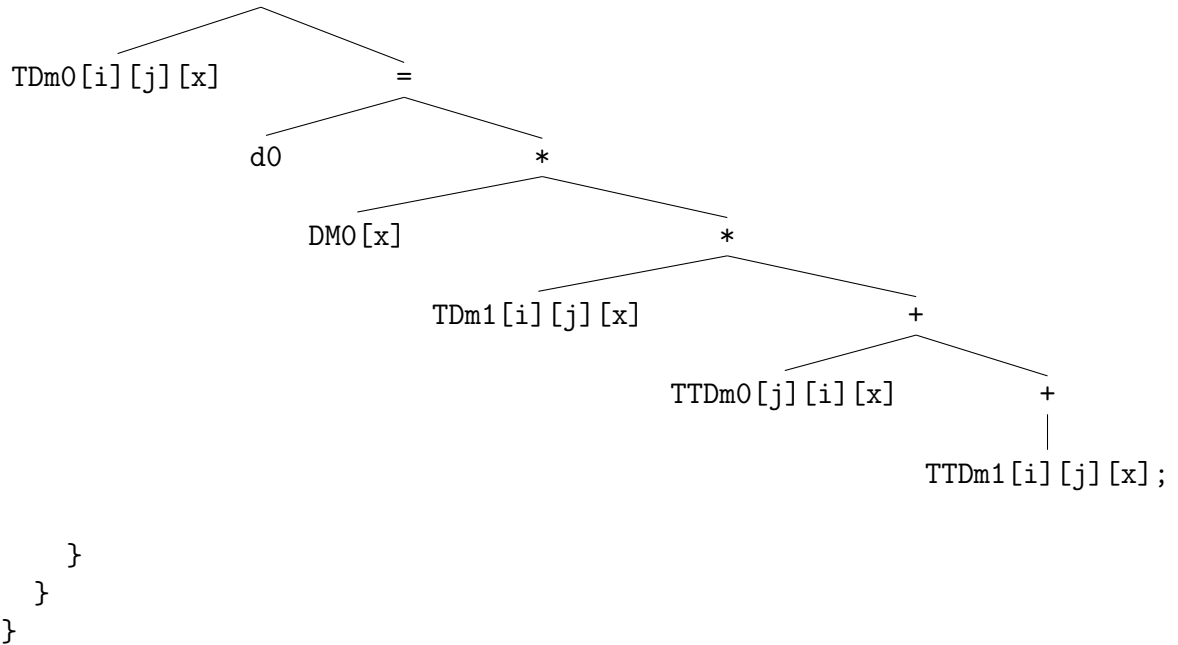

In Figure \ref{fig:kijtreefig} we illustrate the tree structure appropriate
to the operation $\partial_i g_{ij} = -2\alpha K_{ij} + \nabla_i \beta_j + \nabla_j \beta_i$.
The top panel shows the \texttt{TLoops} source-code expression. 
The second and third illustrate the expression tree.
In the second panel that tree is illustrated by a shorthand
representation of the expression template.
\texttt{BOp}, in particular, stands in for the \texttt{iBinaryOp}
introduced in Listing \ref{lst:iBinaryOp} and the surrounding text.

The third panel illustrates the tree recursion performing automatic
C code generation.
First, we generate an appropriate set of \texttt{for} loops from the
index structure of the LHS tensor.
We next iterate through the tree nodes representing
operators, outputting variables that represent concrete data 
(such as \texttt{d0} for \texttt{double}) from the child leafs of each
node.

We now turn to equivalent code output in multiple languages (C and CUDA)
illustrated in Figure \ref{fig:kijtreefig}. Performing
this output using compile-time templates proved cumbersome, since
such template-based code is difficult to write and debug, and is too inflexible
for our diverse goals. We therefore have developed a 
secondary \emph{run-time} representation of the expression tree as
concretely-instantiated C++ classes, 
which works as follows:
\begin{itemize}
\item The abstract \texttt{class TExpressionLeafBase} represents a leaf in the
      expression tree, with concrete derived classes for each type of leaf, such
      as \texttt{double}, \texttt{DataMesh}, or indexed \texttt{Tensor<DataMesh>}.
\item The abstract \texttt{class TExpressionOperatorBase} represents operators,
      with one concrete derived class for each ($+$, $-$, $\texttt{sqrt}$, etc).
\item \texttt{class TExpressionNode} represents a node in the expression, which may be a leaf or an operator. 
      This is the class which forms the actual tree structure, and which handles recursion. It
      holds pointers to any child \texttt{TExpressionNode}s, as well as a pointer to the\\
      \texttt{TExpressionLeafBase*}, if a leaf, or \texttt{TExpressionOperatorBase*}, if an operator.
\end{itemize}

With the above class-reprentation of an expression in hand, we are now ready to
output actual code. Conceptually, each class will trigger output of whatever
code-fragment it represents: 
\begin{itemize}
\item A \texttt{TExpressionOperatorBase} will have member functions to output the
      string representation (`\texttt{+}', `\texttt{sqrt}', etc.). These member
      functions will place the operants in the right places, e.g. on either side
      of binary operators like `\texttt{+}', or within the parentheses of
      unary operators like `\texttt{sqrt()}'.
\item A \texttt{TExpressionLeafBase} has member functions to output any
         code fragments directly involving the operand.  For example, the
         member function \texttt{std::string VarDeclaration();}, which outputs variable
         declarations for the C-style code, outputs
         \texttt{const double d0;}, in the case of the first \texttt{double} on the right hand side,
         or \texttt{const double* TDm1[3];}, in the case of the second \texttt{Tensor} on
         the right hand side, with a rank of 1 and dimension of 3.
\item  A \texttt{TExpressionNode} will have member functions \texttt{PrintExpression}
          and \texttt{PrintCUDAExpression}. In the case of a leaf, these call the appropriate code
          output frunction from \texttt{TExpressionLeafBase}. In the case of an operator, they
          call the output functions of the associated \texttt{TExpressionOperatorBase} along with
          those of the next child \texttt{TExpressionNode}, with parentheses formatted appropriately
          depending on whether the operator is unary or binary.
          \texttt{class TExpressionLeafBase} or 
\end{itemize}

So far, we have described the structure and operation of a fully initialized 
\texttt{TExpressionTree}, and how such a tree yields the desired output code.
We now turn to the construction of these \texttt{TExpressionTree}s. First,
we must interface the templated representation of an expression
(\texttt{iBinaryOp<L, Op, R>}) with this new C++ class-based representation.
To do so, we proceed as follows:
\begin{itemize}
	\item We define a set of C++ functions that are templated on the
              expression-template representation. These functions call one another
              recursively, and in this way they recurse through the expression-template
              representation in the appropriate order, returning at each stage
              the relevant part of the class-based representation, i.e. the
              correct \texttt{TExpressionNode}s.
        \item We further define a templated wrapper class 
	      \texttt{ConcreteTExpressionTreeHouse} which constructs the class-representation
              and which provides convenient functions to interact with it. This class
              is derived from an abstract base-class \texttt{TExpressionTreeHouse},
              which hides the type of the concrete expression-template. Thus, having
              a pointer to \texttt{TExpressionTreeHouse}, surrounding code can
              interact with the expression in a type-agnostic way, making it 
              possible to loop over different expressions and output code.
\end{itemize}

At this stage, we are faced with the task of constructing one 
\texttt{ConcreteTExpressionTreeHouse} for each distinct expression-type in 
\texttt{SpEC}, and collecting pointers to the \texttt{TExpressionTreeHouseBase} in
one large list, so that we can iterate over the tree. For this, we utilize the 
existing code in \texttt{SpEC} named \texttt{Factory}, which implements the 
\emph{factory} design pattern \cite{gamma1994design}.

Conceptually, for each abstract base-class in \texttt{SpEC}, there is a 
database called \texttt{Factory} within which each concrete derived class
\emph{registers} itself, providing an ID-string along with a pointer to 
a function that creates the concrete derived class. After
registration, \texttt{Factory} can then be passed ID-strings, causing it
to call the relevant create-function and to return a pointer to the
newly created instance of the polymorphic class.  

We use the \texttt{SpEC}-\texttt{Factory} as follows. When \texttt{SpEC} is
compiled with the flag $\mathtt{-DCODE_WRITER}$, each \texttt{TLoopApply}-template
function - the function which triggers evaluation of the expression template,
and which is thus itself uniquely templated on the expression - 
activates extra code that defines a static variable
\begin{lstlisting}[language=C++]
static bool registered_... = Factory::Register(options);
\end{lstlisting}
In the above, \texttt{...} represents a unique string constructed from the 
complete template type by recursive calls to a function \texttt{TNameHelper}
mapping expression-template types to string fragments. 

During standard execution,
these extra variables have no effect. They instead become important when linked
into the \texttt{CodeWriter} executable. Upon initialization of the static 
variables at the start of \texttt{CodeWriter} execution, each registered 
variable causes the relevant call to \texttt{Factory::Register}, thus 
building a database containing all expression-templates occuring within
the object files. Each entry in this database will now be represented by a concrete
derived class of the CodeWriter base-class, making the list of expression templates
available at runtime to the executable. At that point \texttt{CodeWriter} 
iterates through all these concrete derived classes, creates one instance
of the expression tree class-representation from each, and calls the
relevant member functions to output C and CUDA code.

\begin{lstlisting}[caption={\label{lst:CodeWriterMainLoop}
      Illustration of main CodeWriter loop.}, language=C++]
void CodeWriter::Write(){
	int fnumber=0;
	const std::list<std::string> Exps
 	  = Factory::RegisteredClassIDs<TExpressionTreeHouseBase>(); //*
	for (auto ExpTag:Exps){
		++fnumber;
		TExpressionTreeHouseBase* TreeHouse_ptr = 
                  TExpressionTreeHouseBase::CreateDerivedClass(ExpTag); //**
		WriteEntry(TreeHouse_ptr, fnumber);
		delete TreeHouse_ptr;
	}
}
\end{lstlisting}

Listing \ref{lst:CodeWriterMainLoop} illustrates the iteration-through-templates
procedure performed by \texttt{CodeWriter}. In that Listing, the line 
marked \texttt{//*} retrieves from the \texttt{Factory} 
a list of all possible ID-strings which represent derived classes
from \texttt{TExpressionTreeHouseBase}, and thus which represent expression templates.
\texttt{CodeWriter} then iterates through that list and,
in the line marked \texttt{//**}, constructs a concrete instance
of \texttt{ConcreteTExpressionTreeHouse} for each expression.

\texttt{CodeWriter} outputs into three files. The first contains functions 
whose arguments are templated on the appropriate
\texttt{TLoops} expression template. These functions route to either the 
CUDA or the CPU code depending on which of 
\texttt{AccelCPU} or \texttt{AccelCUDA} are defined. The second extracts the 
actual arrays of pointers from the \texttt{iBinaryOp}'s passed to the function 
and passes these on. In the CUDA case this array of pointers must be
copied to the GPU via an API call, which can be a significant extra expense. 
To avoid this we make the copy only once,
repeating only if the structure of the \texttt{Tensor} changes. The third file 
contains the actual functions and, in the CUDA
case, tuning arguments for the kernels, which are chosen based on the tensor 
structure of the left-hand side (see Section \ref{Benchmarks} for details).

\section{Design Considerations of Automatically Generated CUDA Code}
\label{CUDAdesign}
Section \ref{AutoGen} detailed the tools we have developed to output
automatically generated C and CUDA code to perform \texttt{TLoops}
operations. In this section we describe the structure of that code with
an eye to its performance.

First, let us give some examples of C-style code generation.
Consider, for example, the TLoops expression
\begin{lstlisting}[caption={\label{lst:TLoopsExample} TLoops rank 2 contraction.}, language=C++]
C(Sym<0,1>(), a_, b_) = Sum(c_, A(a_, c_) * B(c_,b_));
\end{lstlisting}
which symmetrically contracts the rank-2 tensors \texttt{A} and \texttt{B} over their shared
index $\mathtt{c\_}$. 
\texttt{CodeWriter} generates the following C-style code from Listing \ref{lst:TLoopsExample}:
\begin{lstlisting}[caption={\label{lst:cloop}, C-style code corresponding to Listing \ref{lst:TLoopsExample}.}, language=C]
for(int b=0; b<4; ++j) {
  for(int a=b; a<4; ++a) {  \\*
    for(int x=0; x<N; ++x){
      double sum=0;
      for(int c=0; c<4; ++c){
        sum+=A[a][c][x]*B[c][b][x];
      }
      C[a][b][x] = sum;
    }
  }
}
\end{lstlisting}

Recall \texttt{N} is the spatial gridsize.
In Listing \ref{lst:cloop}, \texttt{a} in the \texttt{for} loop marked
\texttt{\\*} is initialized to \texttt{b} rather than to \texttt{0}, due
to the  \texttt{Sym<0,1>()} flag in Listing \ref{lst:TLoopsExample}.
The \texttt{for} loops run up to 4 due to the use of \texttt{a\_, b\_}$\ldots$ rather
than \texttt{i\_, j\_}$\ldots$ indices, which would generate loops running to 3.
Indexing may be further controlled by `fixing' indices (e.g. by specifying
an integer value, such as \texttt{1}, instead of an index such as \texttt{i\_}),
or by specifying index ``offsets" such as \texttt{i\_+1}. Thus, the expression
\begin{lstlisting}[caption={\label{lst:TLoopsExample2} TLoops rank 2 contraction
demonstrating fixed and offset indices.}, language=C++]
D(Sym<0,1>(), i_, 0) = Sum(c_, E(i_+1, c_) * F(c_, 0));
\end{lstlisting}
generates the following C-style code
\begin{lstlisting}[caption={\label{lst:cloop2} C-style code corresponding to Listing \ref{lst:TLoopsExample2}.}, language=C]
for(int i=0; i<3; ++i){
  for(int x=0; x<N; ++x){
    double sum=0;
    for(int c=0; c<4; ++c){
      sum+=E[i+1][c][x]*F[c][0][x];
    }
    D[i][0][x] = sum;
  }
}
\end{lstlisting}
note that in this case the \texttt{Sym<0,1>()} flag has no effect.

Let us now demonstrate our automated CUDA code, starting with a 
simplified sample generated from the expression
\label{lst:prototype}
\begin{lstlisting}[caption={\label{lst:TLoopsExampleNoSym} TLoops rank 2 contraction.}, language=C++]
C(a_, b_) = Sum(c_, A(a_, c_) * B(c_,b_));
\end{lstlisting}
This differs from Listing \ref{lst:TLoopsExample} only in that the symmetry
flag has been removed. In CUDA, instructions to the GPU are collected into
function-like entities called \emph{kernel}s. The kernel generated
from Listing \ref{lst:TLoopsExampleNoSym} closely resembles the following: 
\begin{lstlisting}[caption={\label{lst:CUDAkernel} CUDA kernel corresponding to Listing \ref{lst:TLoopsExampleNoSym}.}, language=C]
__global__ void g_0001(const int N, double** TDm00, 
  const double** TDm01, const double** TDm02){
    const int a = threadIdx.y;
    const int b = blockIdx.y;
    const int x = blockIdx.x*blockDim.x + threadIdx.x;
    if ((x<N)&&(a<4)&&(b<4))
      TDm00[a+4*b][x]=TDm01[a+4*0][x]*TDm02[0+4*b][x] + 
                      TDm01[a+4*1][x]*TDm02[1+4*b][x] + 
                      TDm01[a+4*2][x]*TDm02[2+4*b][x] + 
                      TDm01[a+4*3][x]*TDm02[3+4*b][x]; 
}

\end{lstlisting}
In the real code we use \texttt{\_\_restrict\_\_} flags on the pointer arguments for
performance reasons.
On the GPU, it is advantageous to store the tensor indices in a single, flattened
array, since pointer indirections are relatively expensive. Similarly, 
unrolling expressions which on the CPU would have appeared as
\texttt{for} loops prevents unnecessary serialization. 

The variables \texttt{threadIdx.y}, \texttt{blockIdx.y}, etc, are used by CUDA
to manage parallel data access. In CUDA, computations are abstracted as a 
three-dimensional \emph{grid} of \emph{blocks}, in
turn composed of \emph{threads}. Each thread represents a discrete computational
process that will execute the instructions in the kernel. While differing
threads issue the same instructions, they will normally do so upon differing
data, since they may address memory using their unique block and thread
indices (\texttt{threadIdx.y}, etc.). \texttt{for} loops are generally
replaced with \texttt{if} statements such as that in 
Listing \ref{lst:CUDAkernel}, which ensures that no thread make an out-of-bounds array access.

Physically, GPU 
resources are divided into \emph{streaming multiprocessors} (SMs) composed of tightly
coupled processing cores. Cores in a given SM execute instructions in lockstep,
and share certain memory resources with one another besides the global
memory accessible to the entire GPU. Dividing threads into blocks, which 
are always local to a particular SM, enables such resources to be safely utilized.
Although we make no use of such resources, the blocksize is nevertheless important
for us, since a single SM may operate upon only a certain number
of blocks at one time. The SM hides latency by switching between those blocks
when one stalls (for example because of a data dependency). A poor blocksize
choice can result in low ``occupancy", which impairs this ability, since there
are insufficient blocks to switch to.

For this and other reasons, it is important to appropriately ``tune" the 
kernel launch, via
choice of the number of blocks in the grid (\texttt{nblocks}), 
and the number of threads in each block (\texttt{blocksize}).
Those arguments are in the case of Listing \ref{lst:TLoopsExampleNoSym}
fixed by the corresponding ``wrapper" code:
\begin{lstlisting}[caption={\label{lst:CUDAwrapper} CUDA wrapper corresponding to Listing \ref{lst:TLoopsExampleNoSym}.}, language=C]
void CUDAWrapper_g_0001(const int N, double** TDm00, 
  const double** TDm01, const double** TDm02){
    const int blocksize_x = 64;
    const int nblocks_x = sz/blocksize_x + (sz%blocksize_x == 0?0:1);
    const int blocksize_y = 4;
    const int nblocks_y = 4;
    const int blocksize_z = 1;
    const int nblocks_z = 1;
    const dim3 blocksize(blocksize_x, blocksize_y, blocksize_z);
    const dim3 nblocks(nblocks_x, nblocks_y, nblocks_z);
    g_0001<<<nblocks,blocksize>>>(sz, TDm00, TDm01, TDm02);
}
\end{lstlisting}

We expose the parallelism of the $N$ data-independent spatial gridpoints by 
devoting the entire logical $x$ dimension of the CUDA grid to them. We then
use the four remaining thread addresses to parallelize over the indices of the
left-hand side tensor, since the corresponding arrays are also data-independent. 
In principle,
we could implement the Sum operator as a parallel reduction as well,
since this dramatically complicates automatic code generation and offers
no advantage at dimensions $3$ or $4$, we instead perform them in serial, 
as demonstrated in Listing \ref{lst:CUDAkernel}.

There are two cases in which we cannot parallelize over all the LHS
components. The first is that of a symmetry. For example, 
Listing \ref{lst:TLoopsExample}, which has a symmetry between the 
\texttt{a} and \texttt{b} indices, generates the following kernel:
\begin{lstlisting}[caption={\label{lst:CUDAkernelSym} CUDA kernel corresponding to Listing \ref{lst:TLoopsExample}.}, language=C]
__global__ void g_0001(const int N, double** TDm00, 
  const double** TDm01, const double** TDm02){
    const int b = threadIdx.y;
    const int x = blockIdx.x*blockDim.x + threadIdx.x;
    if ((x<N)&&(b<4)){
      for(int a=b; a<4; ++a){ //*
        TDm00[a+4*b][x]=TDm01[a+4*0][x]*TDm02[0+4*b][x] + 
                        TDm01[a+4*1][x]*TDm02[1+4*b][x] + 
                        TDm01[a+4*2][x]*TDm02[2+4*b][x] + 
                        TDm01[a+4*3][x]*TDm02[3+4*b][x]; 
      }
   }
}

\end{lstlisting}
Because the number of passes through the \texttt{for} loop in marked by
\texttt{//*} in Listing \ref{lst:CUDAkernelSym} depends on the value of
\texttt{b}, parallelization of that loop would require different 
CUDA threads within a block to execute different instructions. Since 
all the threads in a particular SM share the same control circuitry, 
however, this is not possible. CUDA deals with this by having
SMs that encounter so-called ``divergent execution paths" run each one
in serial. We avoid this by serializing explicitly.

The second case we cannot parallize is the unusual one of an LHS tensor
of rank greater than four. This exhausts the number of independent 
CUDA thread addresses, and so we must serialize the extra indices.

\begin{table}
\begin{center}
\begin{tabular}{ |c|c|c|c|c|}
  \hline
  & \multicolumn{2}{|c|}{Bandwidth} & Processing Power \\
  \hline
  Device & theoretical, GB/s & measured, GB/s & GFLOP/s  \\
  \hline \hline
  CPU & 42.7 & - & 8.0 \\
  M2090 & 177.6 & 123 & 665.5 \\
  K80 (one card)& 280 & 170 & 932--1456 \\ 
  P100 & 720 & 449 & 4036-4670 \\ 
  \hline
\end{tabular}
\end{center}
  \caption{Performance specifications for our benchmarked processors. ``CPU" 
    refers to a single core of an Intel Xeon (Sandy Bridge) E5-2620. The K80 actually
    contains two separate GPUs (which share memory) on the same card. Using both requires
    similar extra effort as multi-GPU programming generally, so we profile only one throughout. 
    The K80 and P100 are also potentially
    capable of ``GPU Boost", which dynamically adjusts the core clock 
    frequency if it is possible to do so without exceeding thermal and power limits (the CPU
    has similar capabilities). 
    The ``measured" bandwidths
    were obtained by running the CUDA sample program \texttt{bandwidthTest}.
    }
    \label{tab:GPUtable}
\end{table}

We tune our kernels using the following simple rules, designed to 
achieve maximum or high occupancy on all the GPUs in Table \ref{tab:GPUtable}. 
We use (compare
Listing \ref{lst:CUDAwrapper})
\verb|blocksize_x| and \verb|nblocks_x| to parallelize across the spatial
grid, which leaves us four independent parameters with which to parallelize
across LHS tensor indices.  Each one will be used to parallelize a different
index, and will thus be set to either $1$, $3$, or $4$ (either no index, 
or the dimension of the relevant index). Therefore, 
$\texttt{blocksize\_ y}*\texttt{blocksize\_ z}$ will be either $1, 3, 4, 9, 12,$ 
or $16$.

We now must set \texttt{blocksize\_x} in order to control the \emph{total} blocksize.
This must be a multiple of 32, or else cores will be left idle, 
since GPU instructions are issued to groups
of $32$ in lockstep. An optimal total blocksize, allowing each SM to fully
utilize its compute resources, will be one of a few values
that depend on the particular GPU in question.
On the M2090, for example, these are $192$, $256$, $384$, $512$,
and $768$. $256$ and $512$, in particular, 
achieve maximum occupancy across all three cards.
These values can be achieved exactly when 
\texttt{blocksize\_y}*\texttt{blocksize\_z}
is $1$, $4$ or $16$, in which cases we respectively 
set \texttt{blocksize\_x} to $256$, $64$, or $32$. Otherwise,
we set \texttt{blocksize\_x} to $64$ (for $\mathtt{blocksize\_y}*\mathtt{blocksize\_z}=3$) or 
$16$ (for $12$),
which are near-optimal. Note that this algorithm limits the maximum $N$ that our
code can handle to $65355*\mathtt{blocksize\_x}$, which is always in the millions. This limit
could be easily removed by for example serializing over extremely large
grids, but this has not been necessary for our purposes.

Each SM has a single ``register file" of extremely fast RAM used to 
store variables allocated within a kernel. During kernel launch, those
registers are logically allocated to individual threads as necessary. 
If a kernel's register demands are such that running all possible threads
would exhaust the register file, CUDA will restrict the number of blocks
assigned to each SM to compensate, thus lowering occupancy and possibly
affecting the above calculations. In the above calculations we assume this
does not happen. Our benchmarks (c.f. Table \ref{tab:registertable})
show this assumption is usually, but not always, borne out.

The synchronization of \texttt{Tensor}s presents an additional complication
for CUDA which is not present on the CPU. Recall that the individual arrays 
representing components of a \texttt{Tensor} are not contiguous on the CPU, 
since that class does not assume those components have identical memory 
footprints, and in fact permits modification of those footprints after
construction, for example by reshaping the component \texttt{DataMesh}es. 
The CPU \texttt{Tensor}s instead maintains an array of pointers,
each addressing a particular tensor element. 

This design choice is sufficiently inextricable from \texttt{SpEC}
that we must work around it. But the obvious solution of maintaining an equivalent 
array of pointers on the GPU has a significant performance impact if handled 
naiively. The pointer array can change after construction, so we cannot
simply create a GPU equivalent once and assume it will be always correct. 
On the other hand, the high latency of GPU array allocation and synchronization
makes copying a fresh array with each kernel launch unacceptable. 

Instead, 
each \texttt{Tensor} is paired with a set of ``\texttt{GPUPointers}" that store
a copy of the CPU pointer array, the GPU pointer array, and a reference to the relevant
\texttt{Tensor}. When the GPU pointer array is retrieved, we first ensure that
the copies \emph{CPU} array is identical with that actually present in the 
\texttt{Tensor}, synchronizing only if it is not. This keeps the number of necessary
synchronizations to their bare minimum. If the GPU array is never
retrieved we do not create it at all, so that extra overhead is not incurred
if a \texttt{Tensor} never encounters a \texttt{TLoops} kernel.

\section{Benchmarks}
\label{Benchmarks}

\subsection{Methodology}
\label{Methodology}
We now turn to benchmarks of both the automatically-generated code and the
expression-template implementation. Due to the wide range of potential 
expressions, hardware, compilers, and compiler options, it is not possible to do this 
fully comprehensively, but we aim here to give a broad picture of our
code's behaviour.

Our kernels make no explicit attempt to reuse data once loaded, and we therefore
expect them to be bandwidth-bound or latency-bound 
(i.e. the limiting 
factor to their performance is either the memory bandwidth of the device or the
latency between successive instructions). A useful performance metric in this
case is the ``effective bandwidth" BW$_\mathrm{eff}$, which is the ratio between
the amount of information which must be read and written by an operation with
the time $t$ taken to execute that operation in practice. We measure BW$_\mathrm{eff}$
in GB/s. BW$_\mathrm{eff}$
will be maximal for a kernel which simply copies data, and decrease as operations 
spend significant time on computations
or extraneous memory operations. Calling $N$ the spatial
gridsize, $N_e$ the total number of tensor elements involved in the operation
and $N_d$ the number of doubles occuring outside of a \texttt{DataMesh} array, we have

\begin{equation}
  \label{eq:bweff}
  \mathrm{BW}_\mathrm{eff} = 8 \mathrm{bytes} \frac{N_e N + N_d}{t}.
\end{equation}
For example, the expression in Listing \ref{lst:TLoopsExample} has $N_d=0$ and 
$N_e=42$ ($16$ elements each from \texttt{A} and \texttt{B}, but only
$10$ from \texttt{C} due to the symmetry), while the one in 
Listing \ref{lst:TLoopsExample2} has $N_d=0$ and $N_e=19$ ($3$ from \texttt{D},
$4$ from \texttt{F}, and $12$ from \texttt{E}).

We begin with basic operations typical in relativity. We benchmark each such operation
using three-dimensional tensor indices, at three levels of complexity.
These
operations are assignments
\begin{align}
\label{eq:Assign1}
  A_i &= B_i, \\ 
  \label{eq:Assign2}
  A_{ij} &= B_{ij},\\
  \label{eq:Assign3}
  A_{ijk} &= B_{ijk}, 
\end{align}
additions,
\begin{align}
  A_i &= B_i + C_i, \label{eq:Add1} \\
  A_i &= B_i + C_i + D_i , \label{eq:Add2} \\
  A_i &= B_i + C_i + D_i + E_i, \label{eq:Add3} 
\end{align}
outer products,
\begin{align}
  A_{ij} &= B_i C_j,  \label{eq:Mult1} \\
  A_{ijk} &= B_i C_j D_k,  \label{eq:Mult2} \\
  A_{ijkl} &= B_i C_j D_k E_l,  \label{eq:Mult3}
\end{align}
and contractions
\begin{align}
  A_{ijkl} &= B\indices{_i^m}E_{mjkl},  \label{eq:Contract1} \\
  A_{ijkl} &= C\indices{_j^n}B\indices{_i^m}E_{mnkl},  \label{eq:Contract2} \\
  A_{ijkl} &= D\indices{_k^o}C\indices{_j^n}B\indices{_i^m}E_{mnol}. \label{eq:Contract3}
\end{align}
We furthermore benchmark two practical expressions that actually occur in
numerical relativity. The first mixes scalars, tensors, and outer products,
\begin{equation}
  K_{ij} = 2\alpha g_{ij} + \beta_i \beta_j,
  \label{eq:Kij}
\end{equation}
and the second (Eq. \eqref{eq:SpatialChristoffel}) computes the spatial 
Christoffel symbols of the second kind $\Gamma\indices{^i_{jk}}$, thus
also including contractions. 

Finally, we will present under the label ``GH" our \texttt{TLoops} port of the
actual \texttt{SpEC} module which solves the generalized harmonic equations
in the eponymous formulation of relativity theory. Roughly speaking,
this code computes the second time-derivative of the spacetime metric $\psi_{ab}$,
as a function of first and second spatial derivatives. As such, the input
data is primarily the $\psi_{ab}$, its spatial derivatives
$\partial_i \psi_{ab}$ and $\partial_i \partial_j \psi_{ab}$, and its first
time-derivative $\partial_t \psi_{ab}$. In total there are 22
input arrays. The equations involved consist of 25 separate
\texttt{TLoops} expressions, with as many as four LHS indices, and as many
as two contractions on the RHS. Overall, we estimate 381 spatial-gridsize arrays in the numerator
of Eq. \eqref{eq:bweff} for the GH operation, so that GH is about 4-100 times
more bandwidth-intensive than the other benchmarked expressions. This reflects
in the raw execution times: GH takes about 10 times longer to execute than 
the other benchmarks, though the overall execution time is normalized away
by plotting $\mathrm{BW}_\mathrm{eff}$.

\texttt{TLoops} can also handle transcendental functions and many other unary
functions. Such functions occur often enough that automatic code-generation
is warranted. But they only use a marginal fraction of overall runtime,
and so we do not benchmark such functions at present.

We benchmark each expression on the following combination of hardware and 
code-path:
\begin{enumerate}
  \item Automatically generated CUDA code executing on the three NVIDIA GPUs
      in Table \ref{tab:GPUtable}, namely M2090, K80, and P100.
  \item Automatically generated C code executing on the host processor 
    (labeled `AccelCPU').
  \item The \texttt{TLoops} expression templates executing on the host
    processor (labeled `NonAccel').
  \item The original \texttt{SpEC} code without \texttt{TLoops} code-simplifications,
    as an overall baseline (labeled \texttt{SpEC}.
\end{enumerate}

The CPU code was compiled using \texttt{gcc 4.8.1} using the \texttt{-O3}, 
\texttt{-fPIC}, and \texttt{-std=c++11} compiler flags. We also took benchmarks using 
\texttt{intel 15.0.2} and the compiler flags \texttt{-O3 -xHost -fPIC -fp-model precise -std=c++11}.
The Intel code usually gives comparable or worse results to \texttt{gcc} (c.f. Figure \ref{fig:compilerplot}),
and so for visual simplicity only the \texttt{gcc} results are displayed in
Figures \ref{fig:plotone}, \ref{fig:plottwo}, and \ref{fig:componentplot}.
The CPU timings were performed on a single core of an Intel Xeon E5-2620 CPU, 
which has a clock frequency of 2.0 GHz, a theoretical bandwidth of $42.7$ GB/s, 
and a theoretical double-precision processing power of $8.0$ GFLOP/s. 
We sampled gridsizes at multiples of 32 with decreasing resolution at increasing
gridsize.

We have not made a systematic study of CPUs and compiler options, and do not
intend for these results to reflect the potential CPU performance of our code.
Improved results could almost certainly be achieved using a CPU with a higher
clock frequency and e.g. vectorized instructions over multiple cores.
This machine and these compiler options are, however, representative of 
realistic conditions under which \texttt{SpEC} might presently run. In particular,
limitations to parallelism imposed by \texttt{SpEC}'s implementation of 
multidomain pseudospectral methods restrict it to a single CPU core per expression.

The M2090's host processor is the same as used for the CPU test. We compiled the
CPU code in this case using the Intel compiler with the same options as above. 
The K80 and P100 are hosted by somewhat faster POWER8 processors, and the code
in these cases was compiled using xlc 13.1.4 with 
the flags \texttt{-fPIC, -O3, -std=c++11}.
The CPU's performance should not be relevant to the GPU tests. We compiled the
GPU code with CUDA 6.5 on the M2090 using \texttt{-arch=sm\_20}.
On the K80 (\texttt{-arch=sm\_37}) and P100 (\texttt{-arch=sm\_60}) we used
CUDA 8.0.

\subsection{Benchmarks of simple expressions}
\label{Simple}
We are now ready to present benchmarks for the expressions corresponding to
Eqs. \eqref{eq:Assign1}-\eqref{eq:Mult3} on each of the various hardware
and code-path combinations. We execute each benchmark 21 times, discard
the first, and take the median.

\begin{figure}
  \centering
  \includegraphics[width=\textwidth]{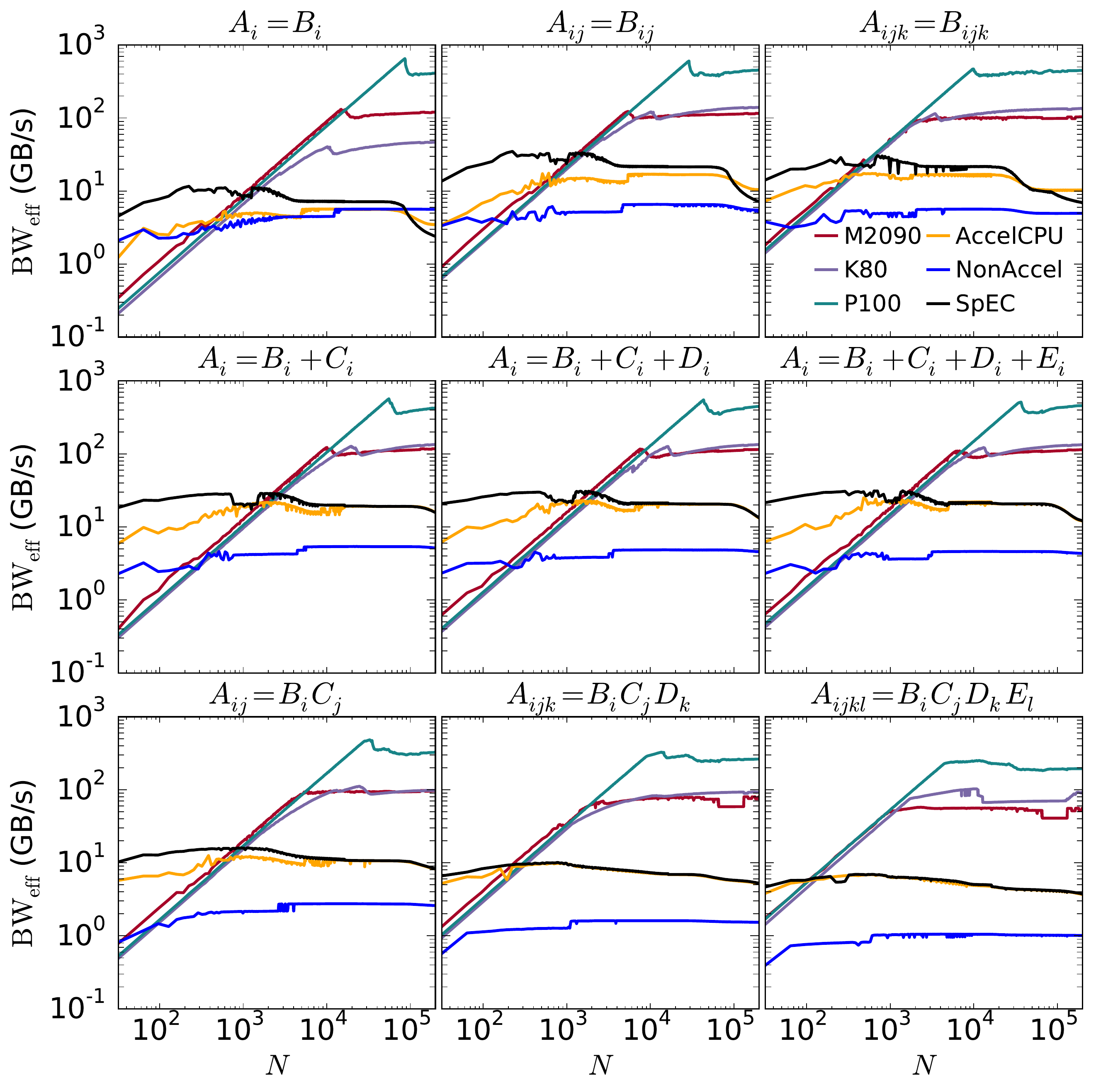}
  \caption{TLoops performance benchmarks of 
  assignment (top row), addition (middle), and outer products (bottom). Operations 
  increase in complexity as panels move from left to right. Each
  panel shows the effective bandwidth of the automatically generated GPU
  code (labelled by the GPU used for the benchmarks), automatically generated
  CPU code (AccelCPU), and expression templates without automatic code generation
  (NonAccel). The black line (SpEC) shows the performance of SpEC without the 
  use of the TLoops package.}
  \label{fig:plotone}
\end{figure}

The results are summarized in Figure \ref{fig:plotone}. Each panel of that
figure correpondes to one particular expression, with the $x$-axis indicated
grid-size, and the $y$-axis indicating performance. We express performances
in terms of effective bandwidth BW$_\mathrm{eff}$, c.f. discussion in 
Section \ref{Methodology} above.

Let us now discuss and interpret Figure \ref{fig:plotone}.
Focusing first on the GPUs, `M2090', `K80', and `P100' all show the same 
overall trends.
Execution time is roughly constant in the gridsize
until occupancy is saturated, after which point it increases linearly. 
Since the number of memory transactions increases
linearly in gridsize throughout, $\mathrm{BW}_\mathrm{eff}$ shows linear increase up
to the point of saturation, after which point it is constant. Since all the
GPUs have essentially the same single-thread performance, they perform essentially
identically until their respective points of saturation. However, newer cards
(especially the P100) can support more parallel threads, resulting in a later
point of saturation with a higher BW$_\mathrm{eff}$.

The saturation gridsize is most importantly determined 
by the left-hand side tensor rank: higher rank tensors
saturate earlier. For example, in Figure \ref{fig:plotone} the 
saturation gridsizes are almost identical for
$A_{ij} = B_{ij}$ compared with $A_{ij} = B_i C_j$, for $A_{ijk} = B_{ijk}$ compared
with $A_{ijk} = B_i C_j D_k$, and for $A_i = B_j$ compared with any of the
addition operations Eqs. \eqref{eq:Add1}-\eqref{eq:Add3}.
This reflects our code's 
parallelism over tensor indices, which is crucial
for achieving good performance at gridsizes on the order of $10^4$. 
The pattern would not persist past rank 4 or
for symmetric indices, since we serialize in these cases.
Post-saturation performance 
is mostly independent of the
operation and usually quite close - slightly beneath a factor of $2$ in the worst case of
$A_{ijkl} = B_i C_j D_k E_l$ -
to the measured bandwidths from \texttt{bandwidthTest}.

In contrast, the three CPU execution-paths do not suffer from high latency,
and so the BW$_\mathrm{eff}$ curves are generally quite flat with respect
to gridsize. Nevertheless, several patterns are visible in the relative exeuction
speed of the three CPU execution paths.
Except sometimes for assignments (Eqs. \eqref{eq:Assign1}-\eqref{eq:Assign3}) 
at very large gridsize, 
the expression-template code 
(\texttt{NonAccel}) usually gives worse performance than either the 
automatically-generated C code (\texttt{AccelCPU}) or that \texttt{SpEC}
without any \texttt{TLoops} simplifications (\texttt{SpEC})
by a factor of between about 3-10. 
These results are roughly in line with
those obtained from Walter Landry's FTensor \cite{FTensorpaper, FTensorbenchmarks},
which is similar to \texttt{TLoops} running in \texttt{NonAccel} mode, and is
presumably due to the compiler being less able to optimize the various templated
expression templates. 

More unexpectedly, \texttt{SpEC} and \texttt{AccelCPU} do not perform identically.
While performance is usually comparable, \texttt{SpEC} is sometimes 
noticeably superior,
especially for less complex operations at
small gridsizes. 
\texttt{AccelCPU} differs from \texttt{SpEC} in two ways. First, the for loops over
tensor indices appear directly in source code using \texttt{SpEC}, whereas
\texttt{AccelCPU} routes through a few extra classes before reaching them.
While we consider it unlikely, the impaired performance for less complex operations 
may be due to some
extra overhead from this routing. Second, \texttt{SpEC} handles the loop over
gridpoints using expression templates, while \texttt{AccelCPU} uses an additional
for loop. This may result in differing behaviour regarding e.g. the creation
of temporaries and the use of cached memory in the machine code.

Now comparing the performance of the GPUs with the CPU executions paths, 
we note that the CPU generally gives better performance at small gridsize, 
but is eventually surpassed by the GPU. This is the expected behaviour:
the CPU has superior single-thread performance, but the GPU has more 
capacity for paralellism.
Compiler optimizations available to the CPU will likely
also result in more efficient reuse of memory than on the GPU. This will make
operations on the CPU less complex, but also make the floating-point performance
of the hardware more relevant. Thus, the CPU has less of an advantage at small gridsize
for more computationally intensive operations. 

\subsection{Benchmarks of more complex expressions}
\label{Complex}
\begin{figure}[t]
  \centering
  \includegraphics[width=\textwidth]{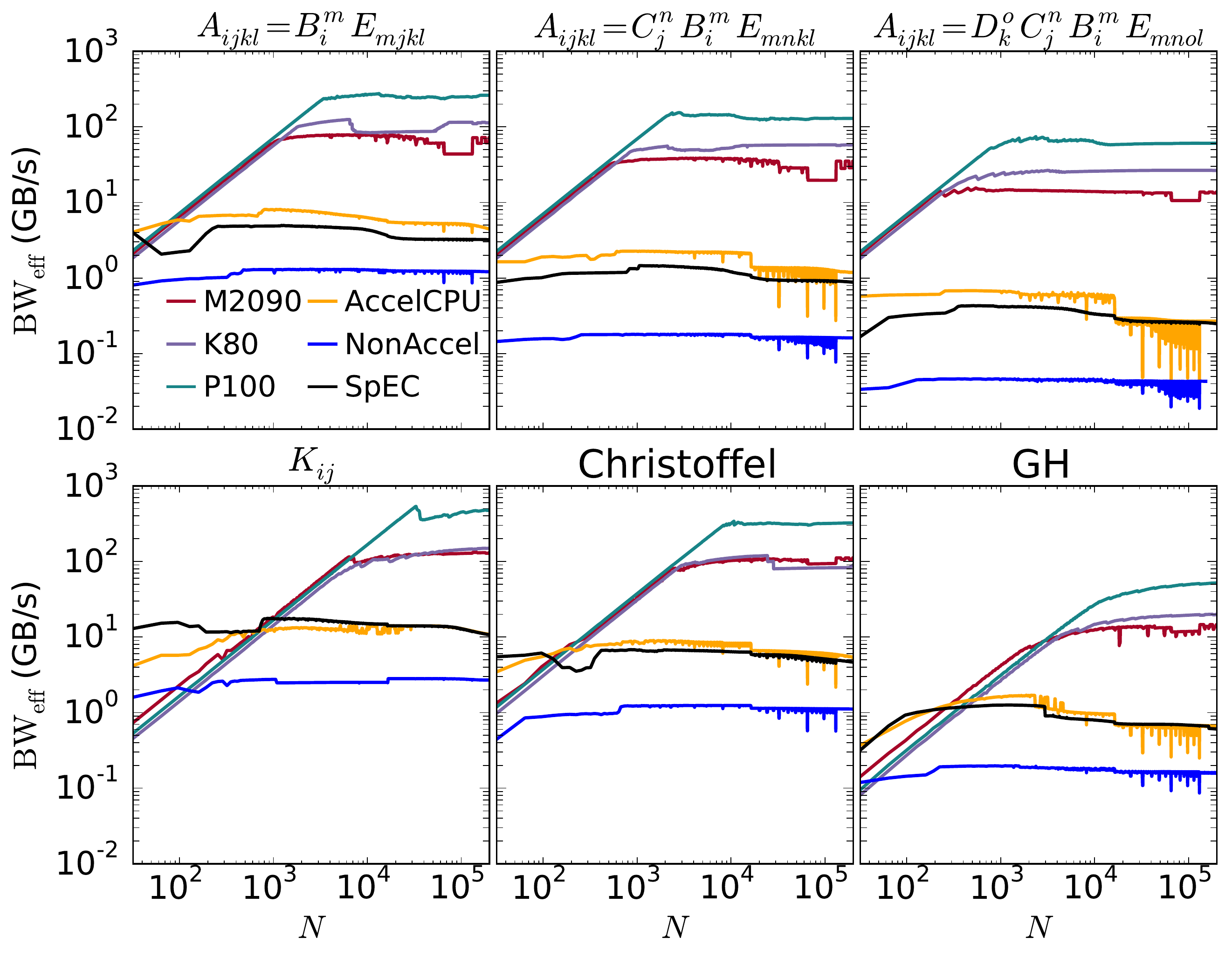}
  \caption{TLoops performance benchmarks of contraction (top row) and of 
  practical numerical relativity operations (bottom). Each panel is formatted
  in the same way as in Figure \ref{fig:plotone}. The leftmost and central
  operations in the bottom panel correspond respectively to Equations \ref{eq:Kij}
  and \ref{eq:SpatialChristoffel}. The rightmost operation, ``GH", shows the
  performance of the entire \texttt{SpEC} module that advances the Einstein 
  equations (in their generalized harmonic formulation) by a timestep.}
  \label{fig:plottwo}
\end{figure}

Let us now turn to the more complex operations corresponding to 
Eqs. \eqref{eq:Contract1}-\eqref{eq:Kij}, Eq. \eqref{eq:SpatialChristoffel}, 
and the GH equations, each described in Section \ref{Methodology}. 
We proceed here as in Section \ref{Simple}, benchmarking six hardware
and code-execution-path combinations as a function of gridsize, with results
presented in Figure \ref{fig:plottwo}. 

Figure \ref{fig:plottwo} shows broadly similar features throughout: GPU performance
increases linearly up to a saturation point and then is constant, while CPU 
performance is nearly flat. In particular, the $K_\mathrm{ij}$ operation
(Eq. \eqref{eq:Kij}, lower left panel of Figure \ref{fig:plottwo}) behaves
essentially identically to Eq. \eqref{eq:Mult2} (lower left panel of
Figure \ref{fig:plotone}), to which it is indeed very similar in form.

The contraction operations 
(Eqs. \eqref{eq:Contract1}, \eqref{eq:Contract2}, and
\eqref{eq:Contract3}) in the top row of Figure \ref{fig:plottwo} show 
some new behaviour. On the CPU, we first of all notice that performance, 
while still independent of gridsize, worsens sharply as we move from left
to right between
panels.
These operations are more strongly
compute-bound than those discussed until now, although the form of our automatically
generated code does not expose this. Since the CPU performs memory operations
relatively better than floating-point computations, its performance degrades
for operations involving more of the latter.

We also notice that \texttt{AccelCPU} gives
better performance than does \texttt{SpEC} for these operations, which is the
opposite behaviour as observed previously. We can only guess at the reason
for this. Perhaps there is more opportunity for compiler optimizations for
operations involving more floating-point operations, but the expression-templates
over gridsize used by \texttt{SpEC} prevent those optimizations from being made.

Turning attention to the GPU curves, we see that the low-gridsize performance
is almost exactly identical between panels. Due to the massive parallelism
it must support, the CUDA compiler cannot make nearly so aggressive optimizations
as can a modern C++ compiler, and so the automatically-generated code 
presumably behaves in the bandwidth-bound manner in which it is written.
The saturation gridsize, however, does change, even though the number of LHS
indices remains constant. Similarly, the post-saturation performance gets 
lower as we move from left to right.

This stems from the fact that the 
\texttt{SpEC} class \texttt{Tensor} is a list of arrays (one array over
the spatial grid per tensor index), rather than a single
contiguous one. Since the tensors are not contiguous each memory access actually
involves two pointer indirections, one each to retrieve the appropriate component
array and spatial gridpoint. For example an instruction such as \texttt{d = A[i][j][x]} must first load
\texttt{A[i]} from global memory, then \texttt{A[i][j]}, then finally \texttt{A[i][j][x]}.
Since the first two loads are not in principle necessary, we do not include them
in the numerator of $\mathrm{BW}_\mathrm{eff}$. Since that numerator therefore 
underestimates the true number of memory transactions our computed $\mathrm{BW}_\mathrm{eff}$
will be correspondingly lower. 

\begin{table}
\begin{center}
\begin{tabular} { |c|c|c|c|c|c|c|}
  \hline
   & \multicolumn{2}{|c|}{M2090} & \multicolumn{2}{|c|}{K80} & \multicolumn{2}{|c|}{P100} \\
  \hline
   Operation & Regs & \% Occ & Regs & \% Occ & Regs & \% Occ \\
  \hline \hline
  \eqref{eq:Assign1} & 10 & 100 & 10 & 93.8 & 12 & 93.8 \\
  \eqref{eq:Assign2} & 10 & 100 & 10 & 93.8 & 12 & 93.8 \\
  \eqref{eq:Assign3} & 10 & 83.3 & 10 & 93.8 & 12 & 93.3 \\
  \hline
  \eqref{eq:Add1} & 14 & 100 & 14 & 93.8 & 14 & 93.8 \\
  \eqref{eq:Add2} & 18 & 100 & 16 & 93.8 & 15 & 93.8 \\
  \eqref{eq:Add3} & 20 & 100 & 21 & 93.8 & 18 & 93.8 \\
  \hline
  \eqref{eq:Mult1} & 14 & 100 & 14 & 93.8 & 14 & 93.8 \\
  \eqref{eq:Mult2} & 18 & 83.3 & 16 & 93.8 & 16 & 93.8 \\
  \eqref{eq:Mult3} & 21 & 83.3 & 21 & 93.8 & 18 & 93.8 \\
  \hline
  \eqref{eq:Contract1} & 28 & 72.9 & 29 & 93.8 & 24 & 93.8 \\
  \eqref{eq:Contract2} & 38 & 41.7 & 42 & 93.8 & 32 & 93.8 \\
  \eqref{eq:Contract3} & 50 & 41.7 & 56 & 93.8 & 48 & 62.5 \\
  \hline
  \eqref{eq:Kij} & 21 & 87.5 & 42 & 93.8 & 32 & 93.8 \\
  \hline
  \eqref{eq:SpatialChristoffel} & 34 & 62.5 & 32 & 93.8 & 32 & 93.8 \\
  \hline
\end{tabular}
\end{center}
\caption{Per-thread register count (Regs) and theoretical occupancy (\% Occ) for benchmarked
  \texttt{TLoops}
  operations on each GPU as measured by the NVIDIA visual profiler. 
On the M2090, K80, and P100,
register use begins to impair occupancy respectively at counts exceeding
21, 64, and 32. The GH operation
is not profiled here since it does not consist of a single kernel. }
\label{tab:registertable}
\end{table}

The extra indirections also result in additional
thread latency, since the thread must stall between the successive loads, and since
the large number of loads may exhaust the SMs memory pipeline. This 
last effect could in principle be alleviated by staggering loads to avoid memory
dependency, but this would complicate automatic code generation considerably.
Future improvements will instead focus on making tensors contiguous.

The extra pointers finally result in
extra thread-local memory allocations, increasing the kernel's per-thread register
count. Each streaming multiprocessor (SM) in a GPU has physically a single register
file that is logically allocated to threads as needed. Each SM is also theoretically
capable of simultaneously executing a certain number of warps, each consisting of 
32 threads, but only if the per-thread register count is small enough that these
warps do not collectively exhuast the register file. 

On the M2090, K80, and P100
respectively, this occurs when the per-thread register count exceeds 21, 64, and
32. The SMs on the K80 and P100 have equally sized register files (of 256kb, compared
to 128kb on the M2090), but
the P100 SMs can also potentially execute more warps, resulting in a lower register
threshold for maximum occupancy. If the limit is saturated by a large threshold,
occupancy will significantly decrease, resulting in an earlier point of saturation
with worse asymptotic performance. We never exceed this threshold on the K80 
(Table \ref{tab:registertable}) but it does sometimes become relevant
for operations involving contractions.

Finally, let us turn attention to the GH operation, in the lower right panel
of Figure \ref{fig:plottwo}. 
On the GPU, the transition from linear to constant performance growth is
not nearly so sharp as for the single-expression operations. 
This presumably reflects an
averaging out between the many saturation points
of the differing expressions within GH.
GH also displays (in all cases) noticeably worse performance
compared to its predecessors in this discussion.
The GH operation consists of many succcessive individual kernels, 
many of which are complex contractions; thus, the above discussion of
contractions applies here as well. On top of this, the many individual 
kernel launches add latency to the GPU execution time.

\subsection{Impact of CPU compiler}
\begin{figure}[t]
  \centering
  \includegraphics[width=\textwidth]{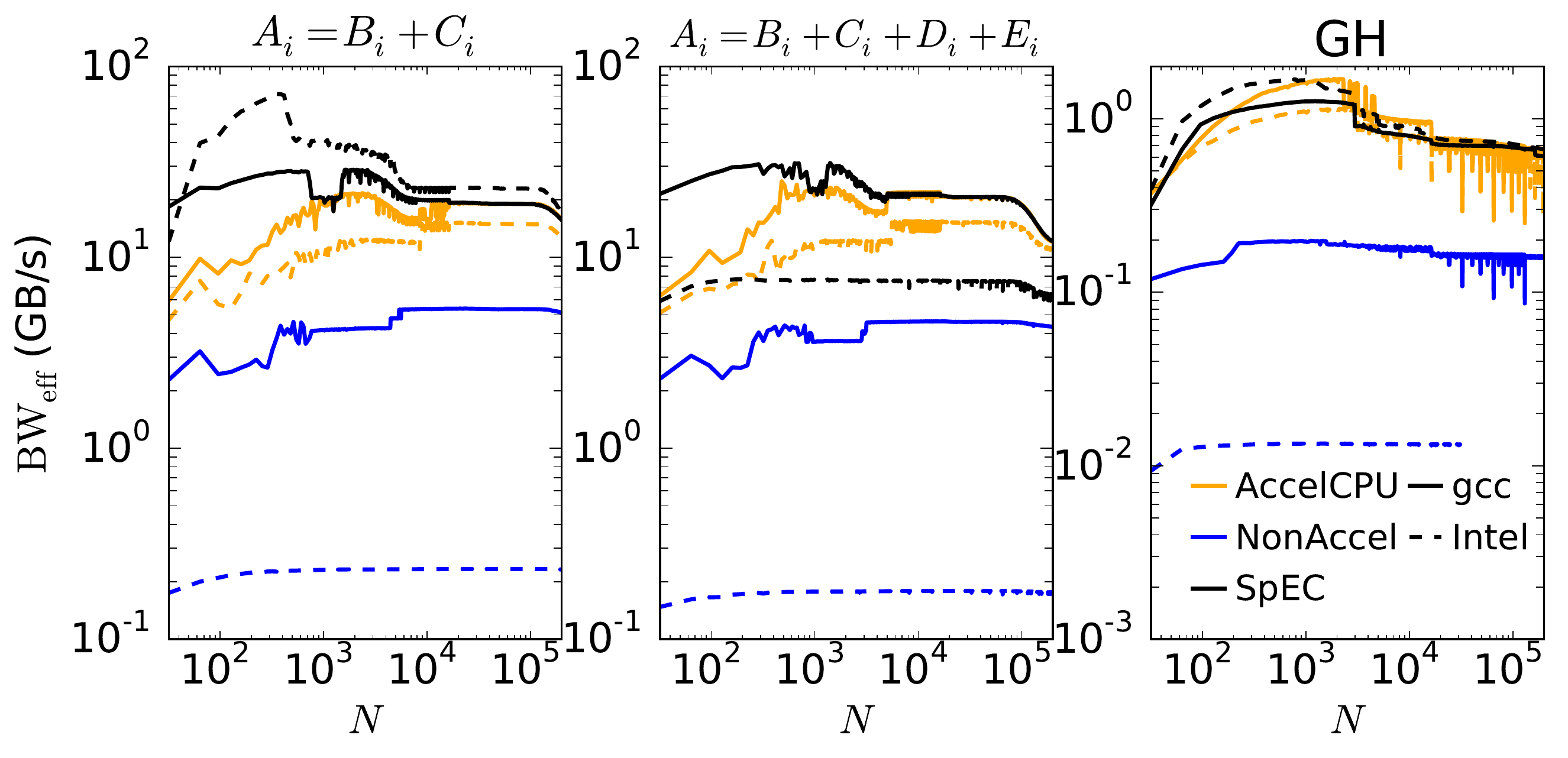}
  \caption{TLoops performance of selected operations illustrating the respective
  performance of the CPU code when compiled using gcc (solid lines) vs Intel
  (dashed) compilers. The detailed version and compiler arguments are 
  given in Section \ref{Benchmarks} of the text. Formatting is otherwise identical
  to Figures \ref{fig:plotone} and \ref{fig:plottwo}.
  }
  \label{fig:compilerplot}
\end{figure}

In Figure \ref{fig:compilerplot}, we show some benchmarks illustrating the
relative performance of gcc vs. Intel compilers operating upon our code. 
Generically, but not always,
gcc gives better performance. The
difference is most stark for the C++11 expression templates of \texttt{NonAccel},
which work over an order of magnitude faster using gcc throughout. gcc
also gets uniformly better performance out of the \texttt{AccelCPU} code,
though the difference is less dramatic. Without \texttt{TLoops} (``\texttt{SpEC}"),
the compilers do behave differently, but their relative performance varies
between operations.

\subsection{Impact of templating over Tensor-indices}
From the perspective of automatic GPU-porting, 
an alternative approach to \texttt{TLoops} 
would be to automatically generate code from \texttt{SpEC}'s existing
spatial-gridpoint expression templates.  
For example, in Listing \ref{lst:Example-SpEC}, one might automatically
generate code only for the interior operation, and not for the full expression
including the \texttt{for} loops over \texttt{i} and \texttt{j}. This
would be much simpler to write, and would require no source code modifications
at all. 

\begin{figure}
  \centering
  \includegraphics[width=\textwidth]{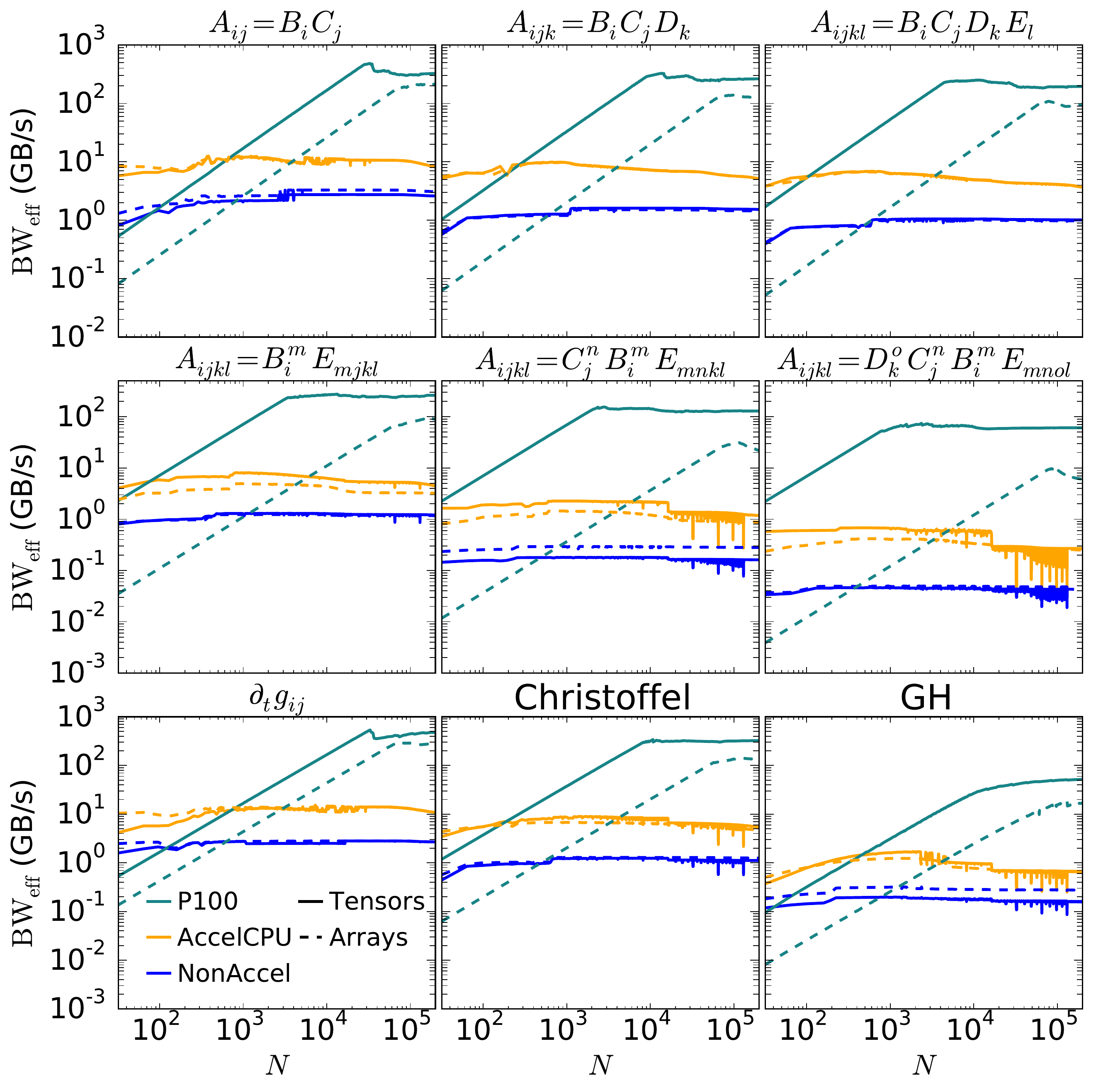}
  \caption{TLoops performance of selected operations showing the performance advantage
  attained by templating over entire tensors (solid lines, ``Tensors" in the legend)
rather than individual component arrays (dashed lines, ``Arrays"). These plots
are otherwise formatted in the same way as in Figures \ref{fig:plotone} and 
\ref{fig:plottwo}. To avoid visual confusion we display results from only one
GPU (the P100), but the qualitative behaviour is the same for each.}
  \label{fig:componentplot}
\end{figure}

Chronologically, this approach was the first we tried. \texttt{TLoops} was motivated
by its strongly negative performance impact on the \texttt{SpEC} code proper.
The performance decrease accounting for this is illustrated in Figure 
\ref{fig:componentplot}. Here, we benchmark various expressions 
using the P100 GPU and the two \texttt{TLoops} CPU 
execution pathways. For the lines marked \texttt{Tensors}, \texttt{TLoops} 
is used to represent the full expression, as in Listing 
\ref{lst:Example-TLoops}. For those marked \texttt{Arrays},
it is used only to represent operations over individual \texttt{Tensor}
elements, which therefore are surrounded by explicit \texttt{for} loops 
in source code, as in Listing \ref{lst:Example-Schematic}.
While the
respective performance of the two strategies is comparable on the CPU, on the 
GPU \texttt{TLoops} expressions are vastly superior, particularly at realistic gridsizes between
about $1000$ and $60000$. 

Templating over tensor indices is advantageous on the GPU for three reasons. 
First, launching a GPU kernel carries an overhead of about 20 $\mu$ s.
In the array loop approach this overhead needs to be paid once per every
free and contracted index in the operation. Automatically ported operations
will usually be small, and in practice launch overhead is very often the 
dominant expense. A \texttt{TLoops} operation, on the other hand, launches only one kernel. 
Second, \texttt{TLoops} operations are paralellized over the left-hand side 
tensor indices as well as the spatial grid, whereas the array loop approach can
parallelize only over the spatial grid. In principle the array loop approach could
achieve some index-level parallelism via concurrent execution of GPU kernels. 
However, the aforementioned launch overhead synchronizes the device, preventing
concurrent execution in practice. 

\section{Conclusion}
We have presented a software package, \texttt{TLoops}, which 
allows tensor-algebraic expressions to compile and execute in C++
code. \texttt{TLoops} can also automatically generate equivalent C++
or CUDA code to these expressions, which can be linked back to a 
second compilation. We have shown this automatically generated code to
give identical or comparable performance compared to the code \texttt{SpEC} uses
by default, and that the CUDA code often outperforms the CPU. Even at only 
moderate gridsizes of a few 1000, the CUDA code
often comes close to the peak (memory-bound) performance of the GPU.

Significant opportunity remains for improvement. The code at present is 
intertwined with the rest of \SpEC. We hope to separate it from the latter
into an independent open-source library. Opportunity for performance improvements
also exists. In particular, we are working on adopting a contiguous \texttt{Tensor}
class within \SpEC. This will allow for simpler, faster automatic code.

We hope the simplifications to coding effort made possible by \texttt{TLoops} may
speed the development of future code, inside and outside of numerical relativity.

\section{Conflicts of Interest}
The authors have no conflicts of interest to report. 

\section{Acknowledgments}
We thank Nils Deppe and Mark Scheel for helpful discussions. 
Calculations were performed with the {\tt SpEC}-code~\cite{SpEC}. 
We gratefully acknowledge support from NSERC of Canada, form the Canada
Research Chairs Program, and from the Canadian Institute for Advanced Research.
Some calculations were performed at the SciNet HPC Consortium~\cite{scinet}.
SciNet is funded by: the Canada Foundation for Innovation (CF) under the auspices of
Compute Canada; the Government of Ontario; Ontario Research Fund (ORF) -- Research Excellence;
and the University of Toronto.

\bibliographystyle{iopart-num}
\bibliography{../Thesis/thesis}

\end{document}